# SoK: Security of Programmable Logic Controllers

Efrén López-Morales
*Texas A&M University-Corpus Christi*

Ulysse Planta
*CISPA Helmholtz Center for Information Security*

Carlos Rubio-Medrano
*Texas A&M University-Corpus Christi*

Ali Abbasi
*CISPA Helmholtz Center for Information Security*

Alvaro A. Cardenas
*University of California, Santa Cruz*

## Abstract

Billions of people rely on essential utility and manufacturing infrastructures such as water treatment plants, energy management, and food production. Our dependence on reliable infrastructures makes them valuable targets for cyberattacks. One of the prime targets for adversaries attacking physical infrastructures are Programmable Logic Controllers (PLCs) because they connect the cyber and physical worlds. In this study, we conduct the first comprehensive systematization of knowledge that explores the security of PLCs: We present an in-depth analysis of PLC attacks and defenses and discover trends in the security of PLCs from the last 17 years of research. We introduce a novel threat taxonomy for PLCs and Industrial Control Systems (ICS). Finally, we identify and point out research gaps that, if left ignored, could lead to new catastrophic attacks against critical infrastructures.

## 1 Introduction

Programmable Logic Controllers, or PLCs, are small rugged computers that have programmable memory to store functions such as timers and logic gates. These functions can control *physical* machines, such as water pumps or centrifuges [36]. PLCs are rather unique systems due to two main characteristics: First, they act as a bridge that connects the cyber and physical worlds; therefore any attack carried out on them can have an immediate effect in the real world. Second, they have opaque, heterogeneous architectures that are different from traditional computer architectures, comprising multiple firmware, which makes them difficult to secure [184].

Before the introduction of PLCs, physical processes were controlled by *relay* panels. Relays are switched either *on* or *off* by an electric current, and thus they can be used in logic circuits. Ladder logic was created as a way to configure relay panels to automatically launch actuation commands based on sensor information. As industrial processes became more complex and varied, relays were not enough to meet the new requirements. An evolution of PLCs, first based on microprocessors, then on networks, and more recently on modern technology paradigms, such as virtualization, has introduced multiple benefits, but at the same time increased the attack surface of these systems.

The first PLC, the *Modicon 084*, was introduced in 1968 [38]. Unlike relays, PLCs could be programmed and reprogrammed to adjust to the process requirements without any physical changes to the control system. A flurry of programming languages emerged, extending ladder logic and introducing new paradigms. Eventually, in 1993, the IEC 61131-3 standard unified these multiple languages into basic standards for PLC programming languages [191].

The first PLCs communicated with the physical world as well as with other computers via analog or serial communications. As computer networks became more reliable and available in the IT world, Ethernet communications were introduced to PLCs. This network connectivity keeps increasing. For example, the Siemens S7-1500 PLC includes up to 4 Ethernet ports, whereas the legacy S7-300 needed an expansion module to support Ethernet connectivity. At the same time, network accessibility now allows remote attackers to deploy classical network attacks against PLCs.

Nowadays, with the advent of Industry 4.0 and IIoT [178], PLCs are going through another paradigm shift bringing even more functionalities like cloud integration, web services, and virtualization. New players like CODESYS and OpenPLC have also entered the market, challenging the long-standing practice of using proprietary hardware and proprietary software products that dominated the PLC industry. These changes pose many open questions about the security of PLCs to the research community.

In order to advance the security of PLCs and the systems they interact with, plenty of research has been produced in the past decades. In particular, research output increased after the term Cyber-Physical System (CPS) was coined in late 2006 by the NSF in the United States [51]. However, the community still lacks an up-to-date general understanding of where the security of PLCs stands and what directions should (or should not) be taken in the future.

To address this challenge, we introduce a comprehensive analysis of the security of PLCs by integrating knowledge from multiple fragmented origins (scientific and grey literature), comprising many existing attack and defense methods in the literature. In addition, we introduce a novel threat model as well as classification and evaluation criteria for summarizing and structuring the existing knowledge about PLC security.

In summary, we make the following contributions:

- We provide a systematization of the literature that consists of **133 papers**, which include **119 attack methods**, and **70 defense methods** (one paper might include one or more attack or defense methods).
- We present a comprehensive taxonomy[1] that integrates the scientific literature with the existing ATT&CK for ICS Matrix in collaboration with MITRE ATT&CK®.
- We identify important PLC security research gaps and discuss future research directions and recommendations.
- We provide **three public tools** to facilitate and foster research and collaboration on this topic. **1)** Our full systematization dataset[2], which other researchers can use to replicate our results and perform their own analyses. **2)** A PLC security artifacts repository[3], which acts as a centralized database for artifacts updated by the community. **3)** A reference graph[4] that enables researchers to further analyze the reviewed literature in this SoK.

## 2 Background

We now introduce some relevant background topics on PLCs that will be leveraged in future sections.

### 2.1 PLC System Environment

PLCs [170] are an essential component of most physical critical infrastructures such as water treatment systems and gas pipelines. PLCs are commonly found in Industrial Control Systems and Supervisory Control and Data Acquisition (SCADA) systems. In the last few years, two new concepts have emerged in the context of Cyber-Physical Systems: Industrial IoT (IIoT) and Induystry 4.0. IIoT is a subset of IoT concerned with connecting industrial assets, e.g., PLCs, with information systems and business processes. Industry 4.0, on the other hand, is a subset of IIoT and refers to the use of Internet technologies to improve production efficiency by employing smart services in smart factories [178]. Figure 12 illustrates the relationship between the above concepts. Furthermore, IIoT and Industry 4.0 have changed PLCs in two ways: First, they have introduced support for modern network protocols, which we discuss further in Sec. 2.3. Second, they have changed the hardware and software architectures to support virtualization and compatibility. We further discuss these topics in Sec. 2.4.

---
[1] https://github.com/efrenlopezm/ics2matrix
[2] https://github.com/efrenlopezm/plc-sok-dataset
[3] https://github.com/efrenlopezm/plcsecurityartifacts
[4] https://www.researchrabbitapp.com/collection/public/E6XRY0186G

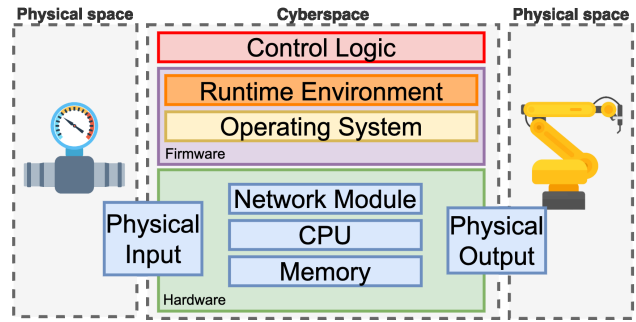

Figure 1: A Generalized PLC Architecture. Based on [36, 69].

PLCs' industrial environment, however, remains largely the same and typically includes the following [185]:

**Actuator.** A hardware component that moves or operates a device in the physical world. Examples of actuators include valves, motors, and piezoelectric actuators.

**Sensor.** A device that generates an electrical analog or digital signal that represents a physical property of a process. Examples include temperature or magnetic field sensors.

**Engineering Station.** A general-purpose computer that is used to write the control logic or ladder logic code for the PLC to execute. It is usually connected to the PLC so that the compiled control logic program can be uploaded.

**Human Machine Interface (HMI).** The hardware or software used to interact with the PLC, e.g, a physical control panel with buttons and lights or a software display.

We further discuss PLCs' underlying architecture and environment in Appendix C.

### 2.2 PLC Basic Components

As shown in Fig. 1, a typical PLC has the following basic components that are susceptible to attacks:

**Control Logic.** A control logic program contains the instructions that the PLC executes to interact with its environment. The IEC 61131-3 standard [191] specifies the syntax and semantics of a unified suite of programming languages for PLCs. Control programs are written in one of these supported languages, e.g., Structured Text, and then compiled into machine code or bytecode. For instance, Siemens PLCs run proprietary MC7 bytecode [144] compiled by the SIMATIC S7 Manager software [175].

**Runtime Environment.** The runtime environment executes the process control code [69] and interacts with the I/O modules. It can be proprietary, e.g., Schneider, or open source like the OpenPLC runtime [19].

**Operating System.** Most PLCs have a Real-Time Operating System (RTOS) [183]. RTOS are operating systems that meet strict processing time requirements and support real-time applications. Vendors support a variety of RTOS in their platforms. For example, Siemens supports the Nucleus RTOS [177] and VxWorks [176].

**Firmware.** The firmware bridges the gap between the PLC hardware and software. While simple PLCs might run applications as bare metal (without an OS) [27, 218], modern PLCs use the firmware under an RTOS. Firmware can be upgraded via SD cards or through a network connection [98].

**CPU.** The CPU interprets the input signals and executes the logic instructions saved in memory. The CPU chassis has slots where other components may be attached, e.g., a network module. It may also include USB ports and SD Card slots.

**Memory Unit.** It stores the program that the CPU will execute along with input data. The memory unit may include different types of memory blocks, which are further discussed in Appendix D.

**Network Module.** Modern PLCs can have one or more ports to communicate with the supervisory control network (regular computers monitoring the process) or the fieldbus (actuators and sensors).

**Physical I/O Modules.** These include input modules with metal pins that receive information (via a voltage or current analog signal) from sensors. The output modules send analog data to actuators such as servo motors.

**PLC Scan Cycle.** It is the cycle in which the PLC reads the sensor inputs, executes the current control logic program and updates the output to the actuators. It is measured in ms and should stay constant. If its time increases, PLCs implement a watchdog timer that sends the PLC to a *Fault* mode [8, 24].

## 2.3 PLC Communication Protocols

As discussed in the introduction, PLCs are becoming more connected via modern communication protocols. Many of these protocols were not designed to include strong security features [101], allowing for the proliferation of vulnerabilities that make the PLCs running them susceptible to attacks [34, 211]. Protocols commonly used in practice include, but are not limited to, the following:

**Fieldbus.** PLCs use Fieldbus [190] protocols to talk to sensors and actuators. Historically these communications were done through serial-based interfaces or analog signals. Sample standards include Profibus, CAN bus, Modbus, and DeviceNet. They all differ in features and implementation, resulting in limited compatibility [67, 112].

**Supervisory Network Protocols.** PLCs communicate with other controllers or classical computers through a variety of proprietary and standardized protocols. These protocols use the IEEE 802.3 Ethernet standard [100] in industrial environments [127]. For example, the EtherNet/IP protocol combines IEEE 802.3 and the TCP/IP Suite [62]. Some are designed to operate on LANs (they do not use IP addresses but only MAC addresses), such as GOOSE, while others use TCP/IP and can be used on LANs and WANs, such as Modbus TCP.

**Industry 4.0 Protocols.** These protocols include support for cloud connectivity, compatibility between devices, and security features. Examples include MQTT (Message Queuing Telemetry Transport) [140] and OPC-UA (Open Platform Communications-Unified Architecture) [197]. Currently, some PLC models support Industry 4.0 protocols [44], whereas older PLCs can be retrofitted to support them [92].

## 2.4 SoftPLCs

The term SoftPLC has been used in the scientific literature since the late 1990s [54, 160, 181]. Although it has not been well defined, generally a SoftPLC is a software runtime environment that executes PLC programs, for example, programs that follow the IEC 61131-3 standard [191]. The runtime is portable and compatible with multiple hardware that can range from microcontrollers to cloud servers [147]. The two major SoftPLC projects are CODESYS [47] and OpenPLC [147]. CODESYS supports approximately a thousand different device types from more than 500 manufacturers [46]. OpenPLC, first proposed in 2014 [22], includes a runtime and an editor, and it is compatible with 18 platforms including Windows and Linux [147]. One of the biggest differences between OpenPLC and CODESYS is that OpenPLC's runtime is open source [19], unlike that of CODESYS. For the remainder of this work, we will refer to traditional PLCs as *HardPLCs*, in order to differentiate them from SoftPLCs.

## 3 Research Questions and Methodology

In this section, we first elaborate on the research questions addressed in this paper, and we then describe the methodology we followed to systematize knowledge by collecting, analyzing, and evaluating works in the literature.

### 3.1 Overview of Research Questions

**RQ-1:** *What are the attack methods against PLCs?* We aim to categorize and analyze attack methodologies targeting PLCs introduced in the literature in the last 17 years.

- **RQ-1.1:** *Which components of the PLC are targeted?* We aim to identify what internal components of a PLC, e.g., the CPU described in Sec. 2.2, are being targeted, in an effort to identify patterns such as the components that have received the most attention in the literature.
- **RQ-1.2:** *How difficult is it to deploy attacks?* We also aim to identify the level of effort required by attackers to successfully carry out attacks against PLCs. In Sec. 4, we elaborate on a threat model and classification criteria, including the potential attack vectors as well as the level of access required by attackers, e.g., Internet access, needed to deploy an attack.
- **RQ-1.3:** *What is the impact of deploying attacks?* Finally, we are interested in identifying the impact that attacks against PLCs may have if they are deployed successfully. This includes the level of disruption achieved, e.g., modifications to control logic programs as shown in Sec. 2.1.

**RQ-2:** *What are the defense methods to protect PLCs?* In addition to attacks, we want to systematize the PLC defenses proposed in the literature in the last 17 years.

- **RQ-2.1:** *Which components of the PLC are protected?* We want to identify which of the components introduced in Sec. 2.2 are the focus of defenses to understand which of them have received less attention and may therefore remain unprotected.
- **RQ-2.2:** *How difficult is it to deploy defenses for PLCs?* We also study the level of effort required to deploy defense mechanisms for PLCs. As shown in Sec. 4.5, this includes the organizational effort from administrators and/or operators to modify and adjust a given PLC/ICS Environment and the performance overhead.
- **RQ-2.3:** *Are there enough defenses addressing reported attacks for PLCs?* Finally, we are interested to know if enough defenses exist in the literature to counteract the attacks we have identified as a part of this work. This way, we aim to identify research gaps and discuss the directions of future work as we describe in Sec. 6 and Sec. 7.

## 3.2 Knowledge Systematization Methodology

To answer the Research Questions raised in Sec 3.1, we use the systematization methodology shown in Fig. 2. First, we perform a systematic literature review. Second, based on the data obtained in the literature review, we perform data analysis and modeling. Finally, we assess the data.

❶ **Systematic Literature Review.** We review both scientific and grey literature to collect PLC attack and defense methods. We also define the final SoK scope based on the specified inclusion criteria.

**Scientific Literature Review.** We consider scientific literature to mean the literature that is based on the scientific method that uses evidence to develop conclusions. It uses previous literature to develop theories and hypotheses while taking care to cite the authors and tools that are used. For the scientific literature review, we carried out the following steps: **1)** Search Querying. We used search queries based on keywords such as *"plc"*, *"ics"*, and *"security"*. The complete list of keywords can be found in Appendix A. **2)** Consulting literature resources. We used these queries to search the selected resources such as Google Scholar and the ACM Digital Library. The complete list of consulted resources can be found in Appendix A. **3)** Applying inclusion criteria. After collecting all these papers, we include them only if they meet the following three criteria. A) The paper must include the term "PLC" or "Programmable Logic Controller". B) It proposes an attack or defense method. C) It includes an experimental evaluation of an attack or defense with at least one PLC. **4)** Applying snowballing. We applied the snowballing technique [209] on all the papers that matched the above criteria to find papers that went unnoticed during the initial search.

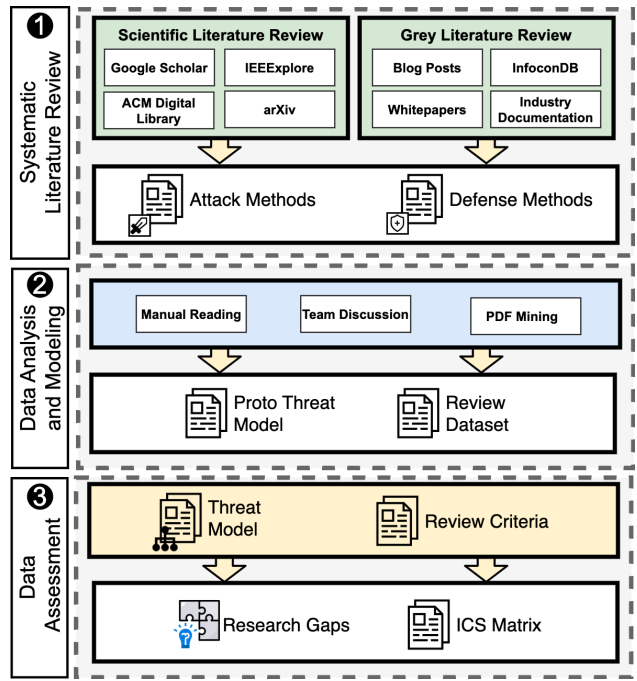

Figure 2: Our systematization methodology. Based on [116].

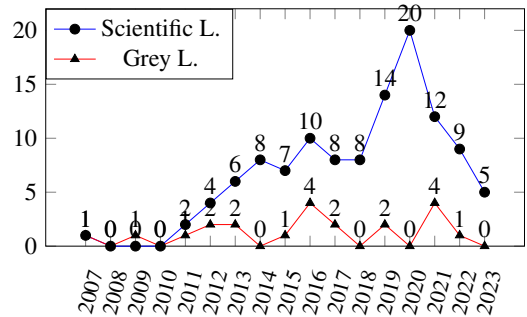

Figure 3: Literature focused on PLC Security per Year.

**Grey Literature Review.** We consider grey literature as literature with limited distribution (i.e., not included in aca-demic publishing libraries). It includes unpublished reports, policy documents, white papers, and technical reports [48]. To perform our grey literature review, we followed the same steps used in the scientific literature review, except that in step 2) (consulting literature resources) we instead searched in the grey literature resources listed in Appendix A.

**Final SoK Scope**. As a result of the scientific and grey literature reviews we collected 133 papers with the earliest being from 2007 and the latest from 2023. Although we searched for papers before 2007, we did not find any that met our criteria. To the best of our knowledge, and based on our exhaustive research, there was no research on designing new attacks or defenses for PLCs before 2007. Fig. 3 shows the final number of scientific and grey papers included in our selection criteria per year of publication, depicting an overall increase in PLC security research over the years, peaking in 2020.

❷ **Data Analysis and Modeling**. In this phase, we first read and analyzed each paper to extract important security-relevant information such as attack vectors, PLC models and manufacturers, and PLC target component. Second, we matched each of the attack and defense methods to their corresponding MITRE technique, subtechnique or mitigation category. The MITRE framework is further discussed in Sec. 4.1. Third, we recorded this information in a spreadsheet. The result of this data analysis and modeling was the identification of the building blocks for a first draft of our threat model depicted in Fig. 4 and the classification criteria discussed in Sec. 4., e.g., the access level and PLC target component.

❸ **Model Assessment**. In this phase, we evaluate the data, threat model, and criteria developed in ❷ to produce the SoK contributions. We use these results to summarize our findings and identify the research gaps discussed in Sec. 6. Finally, we evaluate the results to produce the ICS threat taxonomy discussed in Sec. 4.1.

## 4 Classification and Evaluation Criteria

In this section, we present the criteria to classify and evaluate the works considered for this SoK. These criteria and the symbols introduced alongside them will later be used in Tables 1 and 2. Our first research question RQ-1 is addressed in Sec. 4.2, Sec. 4.3, and Sec. 4.4. Our second research question RQ-2 is addressed in Sec. 4.5, Sec. 4.6, and Sec. 4.7.

### 4.1 ICS Threat Taxonomy

To better classify attack and defense methods in the following sections, we extended the MITRE ATT&CK for ICS Matrix [128] and the Hybrid ATT&CK Matrix [11] to create the ICS$^2$ Matrix, a taxonomy of threats against PLCs and ICS. The taxonomy includes adversary tactics that describe "what" is the adversary's goal and attack techniques that describe "how" the adversary can complete their goal. Additionally, it includes mitigations that prevent a technique from being successfully executed. The ICS$^2$ Matrix incorporates the scientific knowledge accumulated during the past 17 years of PLC security research by adding 6 new attack techniques and 5 new mitigation categories based on the literature reviewed in this SoK. Due to space restrictions, we provide a condensed version of our taxonomy in Fig. 13. The full ICS$^2$ Matrix version is publicly available[5] for the security community to use and extend. Additionally, at the time of writing this paper, we are actively working with MITRE to integrate our research findings within the next version of MITRE ATT&CK for ICS [128] and D3FEND Knowledge Graph [106, 138].

**Matrix Integration.** One of the main limitations of the MITRE ATT&CK for ICS Matrix [128] is that it does not incorporate techniques based on security research findings but instead focuses on techniques based on real-world ICS cyberattacks. For example, the "system firmware" technique [189] includes a reference to Triton discussed in Appendix F. However, it does not include any references to any of the PLC Firmware research listed in Table 2. This was also confirmed by MITRE during one of our communications. Our ICS$^2$ Matrix fills this gap. To integrate our systematization findings with the existing MITRE for ICS Matrix, we needed to add techniques not already covered by MITRE. To develop these new techniques, we used MITRE's official guidelines [137]. This document outlines the information required to create a new MITRE-style technique, for example, a matching "tactic" and "procedure example". In Appendix B we describe in detail how we developed and included one of these techniques.

### 4.2 Attack Classification

Considering RQ-1.1, the following criteria classify attacks according to the taxonomy technique, e.g., Adversary-in-the-Middle, the basic components that are targeted, e.g., the CPU or Memory, and the attack vectors that can be ultimately used against a PLC.

**Target PLC Component.** Each attack method was classified according to the PLC Basic Component that is the primary target, following the description previously shown in Sec. 2.2. Such information was retrieved based on the descriptions explicitly provided by the paper's authors.

**Attack Category.** Each attack method was sorted into one *technique* category, based on our ICS$^2$ Matrix. The technique categories allowed us to determine what kind of defenses might (or might not) counter them.

**Attack Vector.** When launching an attack against a PLC, there might be one or more vectors or paths available to the adversary to deliver the payload. The list of attack vectors we considered are shown in Table 4. Most of the time the attack vector is a vulnerability in the implementation of a network protocol, e.g., S7comm. However, there are other vectors like inserting an SD card into the PLC chassis, for instance, the work by Garcia et al. [75].

---
[5] https://github.com/efrenlopezm/ics2matrix

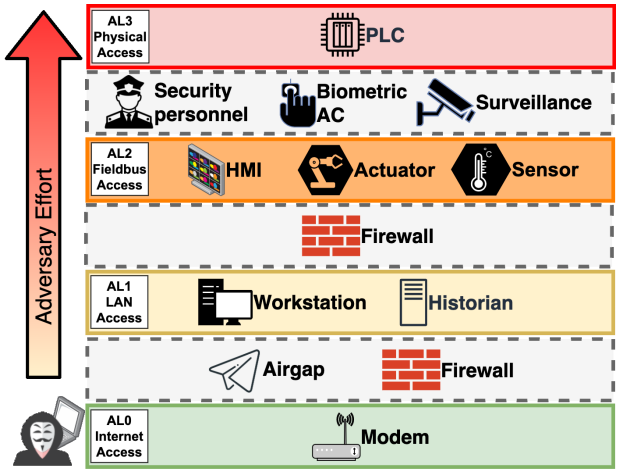

Figure 4: PLC-centered Threat Model. Based on [185].

## 4.3 Attack Complexity

Following RQ-1.2, the criteria shown next are intended to understand, qualify, and quantify how complex it is for an attacker to carry out an attack against a target PLC.

**Environment Knowledge.** This criterion evaluates how much knowledge of the cyber-physical system environment where the PLC lives, e.g., the system topology discussed in Sec. 2.1, is required to launch the attack. An attack may require *zero knowledge* (◯), e.g., a DoS [215] attack targeting the PLC network module only requires an IP address and does not require any environment knowledge. *Partial knowledge* (◐) may require basic information only, e.g., a high-level description of the physical process. CaFDI [132], for example, requires limited local information about the substation configuration. Finally, *extensive knowledge* (●) requires detailed information about the cyber-physical environment. For example, SABOT [131] requires the adversary to encode their understanding of the system behavior into a specification.

**Available Source Code.** This criterion evaluates whether the attack source code is publicly available (●) or not (◯).

**Access Level.** We also consider the level of access an attacker needs to carry out the suggested attacks. Our threat model, shown in Fig. 4, is divided into four distinct *Access Levels* (ALs) labeled from AL0 to AL3.

Adversaries might try to move through the ICS environment in order to reach the PLC and deliver their exploit payload. Overall, our threat model helps us answer two main questions: *1) What access level is required to attempt the attack?*, and *2) What attack surfaces are available to the adversary?*

**AL0: Access to PLC via the Internet (0).** An adversary may be able to access a PLC through an Internet-compatible protocol, e.g., S7comm. This is usually a private network where a remote SCADA computer connects to a PLC via an industrial protocol such as DNP3. Attacks may also be launched if a PLC is publicly exposed on the Internet. At first, this might seem like a ludicrous idea: Why would anybody make a PLC publicly available over the Internet? However, a preliminary Shodan search suggests that the number of publicly accessible PLCs (excluding Shodan-identified honeypots) is over 6,500 as of September 2023[6]

**AL1: Access to Supervisory LAN (1).** This level requires an attacker to have access to the LAN network of the industrial process. This can be achieved if the attacker compromises workstation computers, data historians, or similar operational computers. Some industrial protocols run only on Ethernet and are non-routable (such as GOOSE), so an attacker targeting a vulnerable GOOSE stack on a PLC will have to do so in the same LAN.

**AL2: Access to PLC Fieldbus Network (2).** As mentioned in Sec. 2.3, fieldbus represents the lower level of communications between the PLC and nearby field equipment such as actuators and sensors. An attacker can compromise any of these devices to exploit a vulnerability in the fieldbus code implementation. The attacker can also launch Adversary-in-the-Middle attacks in this network, as fieldbus communications are rarely authenticated in practice.

**AL3: Physical Access to PLC (3).** Lastly, this level assumes the adversary has bypassed environmental and physical protection measures (e.g., locked doors) of the target. Having physical access to the PLC may be the most difficult scenario for an adversary but can still be achieved by malicious insiders or disgruntled employees. An attacker with physical access can use a JTAG interface, as shown in Table 4.

## 4.4 Attack Impact

Following RQ-1.3, the criteria below are intended to understand and quantify the potential impact of a successfully carried out attack, e.g., the *payoff* in return of the effort invested, as discussed in Sec. 4.3.

**Potential Damage.** This criterion evaluates the *immediate* damage inflicted to the physical operation of the ICS. The damage is *limited* (1) if the attack does not change the operation of the process but instead aims to collect information or gain unauthorized access, for example, a password sniffing attack [205]. The potential damage is *substantial* (2) if the attack can stop the industrial process partially or completely, e.g., a DoS attack [146]. The potential damage is *severe* (3) if the attack is able to insidiously control the physical process, e.g., a logic bomb or firmware modification attack, where the attacker can launch arbitrary control commands to the system (similar to Stuxnet, discussed in Appendix F).

**HardPLC Targets.** An attack may affect a *single* PLC model (◯), e.g., the Siemens SIMATIC S7-300, or may affect *multiple* PLC models of the same manufacturer (◐), e.g., the Rockwell's Allen-Bradley 1100 and 1400 models. Also, an attack may affect *multiple* PLC models from two or more manufacturers (●), e.g., Siemens' SIMATIC S7-1500 and Modicon's M221. An attack that is effective against various devices is an indicator of the severity of the problem, while an attack focusing on a single device might be contained.

---
[6] https://www.shodan.io/search?query=plc+-%22792%2F71644%22.

**SoftPLC Targets.** Finally, this criterion indicates if a SoftPLC was used to evaluate the proposed attack. An attack may have been evaluated using *CODESYS* (C), *OpenPLC* (O) or *no SoftPLC* at all (-) (i.e., the attack only focused on HardPLCs).

### 4.5 Defense Classification

After describing our criteria for attacks, we now focus on defenses by looking at RQ-2.1.

**Defense Category.** Each defense method was sorted into one *mitigation* category, based on our ICS$^2$ Matrix. The mitigation categories allowed us to determine what kind of attacks might (or might not) be countered.

**Defense Vector.** The defense vector is the path used in order to stop the attack payload from being delivered. The available defense vectors are defined in our Threat Model for PLCs in Table 4. Most of the time, the attack and defense vectors will match. However, they can differ. For example, *Rajput et al.* use the JTAG interface as a defense vector [157].

### 4.6 Defense Deployability

Following RQ-2.2, the following criteria evaluate how difficult it is to deploy a defense method to protect PLCs.

**PLC Overhead.** This criterion quantifies the overhead that the defense method incurs on the PLC itself, which can be either *zero* (①), *negligible* (②), or *considerable* (③).

**Infrastructure Overhead.** This criterion quantifies the cost that comes with setting up the required infrastructure to implement a defense method. That may involve *no changes* (①), or it may involve *either* an infrastructure change, e.g., setting up new VLANs or an additional hardware component, e.g., bump-in-the-wire (②). Finally, a defense may involve *both* infrastructure changes and hardware components (③).

**Maintenance.** This criterion quantifies the level of post-deployment maintenance required by a given defense method. A defense method may require *no maintenance* (①), *sporadic maintenance* (②), or *constant maintenance* (③).

**Source Code Availability.** This criterion evaluates whether the defense source code is publicly available (●) or not (○).

### 4.7 Defense Robustness

Following RQ-2.3, we also characterize how the defense mitigates attacks.

**Defense Stage.** Defenses can be categorized in three different stages of the security process. 1) *Prevention Defense* (PR) aims to reduce the possibility of an incident *before* a known vulnerability is exploited. 2) *Detection Defense* (DE) aims to identify and alert about a *current or recent* attack. 3) *Recovery Defense* (RE) aims to reduce the damage of an attack *after* it has been carried out successfully.

**Effectiveness.** The defense method's effectiveness is indicated by its accuracy. The accuracy values are taken from the research paper itself. If the paper did not report or specify the accuracy, it is marked as *N/S* (not specified). We selected accuracy because it is the most reported metric across the defense method papers. We recognize that this metric does not work for all defense methods, however, it is the most frequently reported quantitative metric available in the current literature. We further discuss this problem in Sec. 7.

**HardPLC Scope.** This criterion evaluates the number of different PLC models that are protected by the defense. The defense may protect a *single* PLC model (○), e.g., the Siemens SIMATIC S7-300, or it may protect *multiple* PLC models of the same manufacturer (◐), e.g., the Rockwell's Allen-Bradley 1100 and 1400 models. Also, a defense may protect *multiple* PLC models from two or more manufacturers (●), e.g., Siemens' SIMATIC S7-1500 and Modicon's M221.

**SoftPLC Scope.** This criterion indicates if any SoftPLC was used to evaluate a defense, either *CODESYS* (C), *OpenPLC* (O) or *no SoftPLC* at all (-).

## 5 Overview of Attacks

By using the systematization methodology presented in Sec. 3 and the criteria discussed in Sec. 4, we obtained Tables 1, and 2. These tables use the *PLC target component* categorization defined in Fig. 1.

We now illustrate examples of attacks per target component. Due to the large number of papers in our SoK (133 papers), we cannot discuss all of them. To select the papers systematically, we picked the two papers with the most citations from each category. We used Google Scholar to perform the search on June 3, 2023. In the next paragraphs, we discuss the insights provided by these highly-cited papers.

**Attacks that Target the PLC Communications Module.** *Urbina et al.* [198] introduce an AitM attack between the PLC and the field devices (AL2). An interesting observation of that paper is that field networks tend to follow a ring topology rather than the typical star topology of switched Ethernet networks; therefore, an AitM attack does not need to use ARP (Address Resolution Protocol) spoofing or similar techniques to place itself between two devices. It just needs to inject the attack between the two attacked devices. An AitM attacker can then send false sensor readings to the PLC or false control commands to actuators.

*Wardak et al.* [205] introduce another network attack, which focuses on password sniffing on the network interface between a workstation and the PLC (AL1) in Fig. 4. The authors show that several of the connections between the workstations and the PLCs are not encrypted (or authenticated). Consequently, passwords can be sniffed and then used to gain

| Target Component | Attacks | | | | | | | | | Defenses | | | | | | | | | | |
|---|---|---|---|---|---|---|---|---|---|---|---|---|---|---|---|---|---|---|---|---|
| | Method | Category (Technique) | Attack Vector | Complexity | | Source Code | Impact | | | Method | Category (Mitigation) | Defense Vector | Deployability | | | Source Code | Robustness | | | |
| | | | | Access Level | Env Knowledge | | P. Damage | HPLC Targets | SPLC Targets | | | | PLC Overhead | Infra Overhead | Maintenance | | Effectiveness | D. Strategy | HPLC Scope | SPLC Scope |
| Network | Attacks on situational awareness [115] | Adversary-in-the-Middle | P | 1 | ● | O | 2 | O | - | Cross-layer fingerprinting [71] | Network Intrusion Pre. | D | 1 | 1 | 1 | O | 92.8 | DE | O | - |
| | Concealment Attack [61] | | M | 1 | O | ● | 2 | O | - | DFA-based intrusion detection [83] | | M | 1 | 2 | 1 | O | 99.8 | DE | O | - |
| | Controller Eavesdropping [78] | | P | 1 | O | O | 1 | ● | O | Encrypted Traffic IDS [95] | | I | 1 | 2 | 1 | O | N/S | DE | - | - |
| | Controller Packet Tampering [78] | | P | 1 | O | O | 3 | ● | O | IDS for S7 networks [110] | | S7 | 1 | 2 | 1 | O | 99.8 | DE | O | - |
| | False Data Injection Attack [43] | | M | 0 | O | O | 2 | ● | - | Model-based anomaly detection [113] | | S7 | 1 | 2 | 1 | O | 96 | DE | O | - |
| | Man in the middle [25] | | E | 1 | O | O | 1 | O | - | Physical fingerprinting [71] | | D | 1 | 1 | 1 | O | 92.8 | DE | O | - |
| | OPC-UA Rogue Client [10] | | UA | 0 | O | O | 2 | O | - | PLC Watermarking [8] | | E | 1 | 2 | 1 | O | N/S | DE | ● | - |
| | Replay attack [205] | | P | 0 | O | O | 2 | O | - | Embedding Encryption [20] | Encrypt Network Traffic | M | 1 | 1 | 1 | O | N/S | PR | - | O |
| | Replay attack [167] | | F | 0 | O | O | 2 | O | - | Hash Authentication [74] | | F | 1 | 1 | 1 | O | N/S | PR | O | - |
| | SDN Enabled MitM [78] | | P | 1 | O | O | 1 | ● | O | LVST (LabView SSH Tunnel) [153] | | E | 1 | 2 | 1 | O | N/S | PR | ● | - |
| | ISO-TSAP Replay Attack [31] | | P | 0 | O | O | 2 | ◐ | - | AES-256 Implementation [21] | | M | 1 | 1 | 1 | ● | N/S | PR | - | O |
| | Third-party Eavesdropping [78] | | P | 1 | O | O | 1 | ● | O | PLCrypto [212] | | E | 1 | 1 | 1 | ● | N/S | PR | O | - |
| | Change the IP [163] | Denial of Service | E | 0 | O | ● | 2 | O | - | PLC-Sleuth [213] | Network Intrusion Pre. | S7 | 1 | 2 | 1 | O | 100 | DE | O | - |
| | Control engine attack [155] | | M | 0 | O | O | 2 | ● | - | Semantic IDS [89] | | M | 1 | 2 | 1 | O | N/S | DE | - | - |
| | Denial of Service Attack [215] | | P | 0 | O | O | 2 | O | - | Telemetry IDS [154] | | M | 1 | 2 | 1 | O | 99.5 | DE | - | - |
| | DoS clearing Flow Table [78] | | P | 0 | O | O | 2 | O | O | Zeus [90] | | E | 1 | 2 | 2 | O | 98.9 | DE | O | - |
| | Flow Rule Blocking [78] | | P | 0 | O | O | 2 | O | O | Side-channel Anomaly Detection [58] | Out-of-Band Comms Channel | E | 1 | 1 | 2 | O | N/S | DE | O | - |
| | UDP reflect attack [167] | | F | 0 | O | O | 2 | O | - | Anomaly detection [222] | | N | 1 | 2 | 2 | ● | 90 | DE | ● | - |
| | Authentication Bypass [87] | Modify Auth. Process | E | 1 | O | O | 1 | O | - | Arcade.PLC [33] | Control Logic Verification | S7 | 1 | 1 | 1 | O | N/S | PR | O | - |
| | Cryptographic attack [167] | | F | 1 | O | O | 1 | O | - | ShadowPLCs [104] | | S7 | 2 | 1 | 2 | O | 97.3 | DE | O | - |
| | Replay attack [120] | | S7 | 0 | O | O | 1 | ◐ | - | Snapshooter [102] | Network Intrusion Pre. | S | 2 | 2 | 1 | O | N/S | DE | - | O |
| | S7 Authentication Bypass [31] | | P | 0 | O | O | 1 | ◐ | - | Traffic Data Classification [117] | | S7 | 1 | 1 | 1 | O | 99.7 | PR | O | - |
| | Unauthorized password updating [205] | | S7 | 0 | O | O | 1 | O | - | vBump [192] | | G | 1 | 3 | 2 | ● | N/S | PR | O | C |
| | SoMachine Authentication [25] | Hardcoded Credentials | M | 0 | O | O | 1 | O | - | | | | | | | | | | | |
| | Credentials from storage [25] | | E | 0 | O | O | 1 | ◐ | - | | | | | | | | | | | |
| | subverting read/write-protection [25] | | P | 0 | O | O | 1 | O | - | | | | | | | | | | | |
| | subverting write-protection [25] | | P | 0 | O | O | 1 | O | - | | | | | | | | | | | |
| | Command packet flooding [86] | Network Denial of Service | E | 0 | O | O | 1 | O | - | Modbus/TCP Firewall [173] | Filter Net. Traffic | M | 1 | 2 | 1 | O | N/S | PR | O | - |
| | Modbus Flooding Attack [32] | | M | 0 | O | O | 2 | O | - | | | | | | | | | | | |
| | Network Flooding Attack [146] | | M | 0 | O | O | 2 | ● | - | | | | | | | | | | | |
| | UDP flooding Attack [109] | | M | 0 | O | O | 2 | O | - | | | | | | | | | | | |
| | Password reset attack [25] | Brute Force | M | 0 | O | O | 1 | O | - | Shade [217] | Network Intrusion Pre. | E | 2 | 1 | 1 | ● | N/S | DE | ● | - |
| | CLIK password attack [105] | | M | 0 | O | ● | 1 | O | - | Host Anomaly Detecton [69] | | M | 2 | 1 | 1 | O | N/S | DE | ● | - |
| | Dictionary Attack [37] | | B | 0 | O | O | 1 | O | - | Argus [5] | | E | 1 | 3 | 3 | O | N/S | DE | - | - |
| | Dump module code [163] | Data from Local System | E | 0 | O | ● | 1 | O | - | PLC-PROV [9] | Validate Program Inputs | N | 1 | 2 | 2 | O | N/S | DE | - | C |
| | Memory Logic Attack [31] | | P | 0 | O | ● | 1 | O | - | PLCPrint [50] | | S7 | 1 | 1 | 1 | O | 95 | DE | ● | - |
| | Crash 1756-ENBT module [163] | Device Restart/Shutdown | E | 0 | O | ● | 2 | O | - | ABAC model for PLC [85] | Authorization Enforcement | S7 | 1 | 2 | 1 | O | N/S | PR | ● | - |
| | Reset 1756-ENBT module [163] | | E | 0 | O | ● | 2 | O | - | FINS detection rules [73] | Network Allowlists | F | 1 | 2 | 1 | O | N/S | PR | ◐ | - |
| | Leak Modbus data [196] | Exfiltration Side-channel | M | 0 | O | O | 1 | O | C | | | | | | | | | | | |
| | Leak OPC-UA Process Data [196] | | UA | 0 | O | O | 1 | O | C | | | | | | | | | | | |
| | Passive network scanner [206] | Network Conn. Enumeration | M | 1 | O | O | 1 | ● | - | | | | | | | | | | | |
| | Port scanner [111] | | P | 0 | O | ● | 1 | O | C | | | | | | | | | | | |
| | SOCKS Proxy [111] | Connection Proxy | P | 0 | O | O | 1 | O | C | SDN-enabled automatic response [142] | Network Intrusion Pre. | E | 1 | 2 | 1 | ● | N/S | DE | - | - |
| | New ADMIN account [37] | Exp. for Credential Access | B | 0 | O | O | 1 | O | - | WeaselBoard [141] | Exploit Protection | BP | 3 | 3 | 3 | O | N/S | DE | ● | - |
| | Getting a Shell on the PLC [31] | Exp. for Privilege Escalation | P | 0 | O | O | 3 | ◐ | - | Memory Access Taintedness [29] | Exploit Protection | M | - | 3 | 1 | O | N/S | DE | - | - |
| | Password stealing [205] | Network Sniffing | P | 0 | O | O | 1 | O | - | SCADA Protocol Obfuscation [28] | Encrypt Net. Traffic | M | 1 | 1 | 1 | O | N/S | PR | O | - |
| | Wireless Fieldbus MitM [198] | Wireless Sniffing | E | 3 | O | O | 1 | O | - | Secure Logging for ICS [168] | Encrypt Net. Traffic | S | 1 | 3 | 1 | O | N/S | PR | O | - |
| OS | 3S CoDeSys Tools [207] | Exp. for Privilege Escalation | C | 1 | O | ● | 2 | - | C | | | | | | | | | | | |
| | Remote arbitrary code execution [108] | | P | 0 | O | O | 2 | O | - | ECFI [3] | Exploit Protection | RT | 2 | 2 | 1 | O | N/S | PR | O | O |
| | Arbitrary Code Execution [4] | Exp. for Client Execution | R | 3 | O | O | 2 | O | - | | | | | | | | | | | |
| Runtime | Ghost in the PLC [2] | Block I/O Communication | R | 3 | O | O | 3 | ● | C | GhostBuster [1] | Exploit Protection | RT | 2 | 1 | 1 | O | N/S | DE | O | C |
| | Unauthenticated file read/write [82] | Data from Local System | C | 1 | O | O | 2 | O | C | | | | | | | | | | | |
| | XML bomb [82] | Modify Parameter | C | 1 | O | O | 2 | O | C | | | | | | | | | | | |

Table 1: A Summary of PLC Attack and Defense Methods (Network, OS and Runtime Components).
**D**=DNP3, **G**=GOOSE, **S7**=S7COMM, **I**=IEC 104, **UA**=OPC-UA, **P**=Profinet, **E**=EtherNet/IP, **M**=Modbus TCP, **F**=FINS, **S**=Syslog, **B**=Beckhoff, **BP**=Backpane, **R**=SoC Register, **RT**=Runtime, **O**=OpenPLC, **C**=CODESYS / CODESYS upload protocol, **N**=Not Specified, **DE**=Detection, **PR**=Prevention, **RE**=Recovery

| Target Component | Attacks | | Attack Vector | Complexity | | Source Code | Impact | | | Defenses | | Defense Vector | Deployability | | | Source Code | Robustness | | | |
| --- | --- | --- | --- | --- | --- | --- | --- | --- | --- | --- | --- | --- | --- | --- | --- | --- | --- | --- | --- | --- |
| | Method | Category (Technique) | | Access Level | Env Knowledge | | P. Damage | HPLC Targets | SPLC Targets | Method | Category (Mitigation) | | PLC Overhead | Infra Overhead | Maintenance | | Effectiveness | D. Strategy | HPLC Scope | SPLC Scope |
| Control Logic | Advanced Stealthy Injection Attack [17] | Modify Control Logic | P | 0 | 0 | ● | 3 | ○ | - | Safety Verification [77] | Control Logic Verification | S7 | 2 | 2 | 1 | ○ | N/S | DE | ○ | - |
| | CLIK [105] | | SM | 1 | 0 | ● | 3 | ○ | - | CPLCD [214] | | S7 | 1 | 1 | 1 | ○ | N/S | DE | ○ | - |
| | Data Execution Attack [216] | | M | 0 | 0 | ○ | 2 | ○ | - | Trusted Safety Verifier [133] | | N | 1 | 2 | 1 | ○ | N/S | DE | - | - |
| | Fragmentation and Noise Padding [216] | | M | 0 | 0 | ● | 2 | ● | - | Detection Manipulation [119] | | X | 1 | 1 | 1 | ○ | N/S | DE | ○ | - |
| | Ladder Logic Bomb [84] | | U | 3 | 0 | ○ | 3 | ○ | - | PLCloud [172] | | S7 | 1 | 3 | 3 | ○ | N/S | RE | ○ | - |
| | Time-of-Day (TOD) interrupt attack [16] | | S7 | 0 | 0 | ○ | 3 | ○ | - | PLC Guard [125] | | S7 | 1 | 2 | 2 | ○ | N/S | PR | ○ | - |
| | Control logic injection attack [15] | Modify Program | X | 0 | 0 | ○ | 3 | ○ | ○ | PLC-VBS [124] | Vulnerability Scanning | S7 | 1 | 1 | 1 | ○ | N/S | PR | ○ | - |
| | Dynamic Malware Payloads [129] | | N | 1 | 0 | ○ | 3 | ○ | - | ICSPatch [143] | Update Control Logic | N | 1 | 2 | 1 | ● | 100 | PR | ● | C |
| | Executing arbitrary ladder logic [82] | | N | 0 | 0 | ○ | 2 | ○ | - | D-Box [135] | Limit Access to MCU R. | X | 2 | 1 | 1 | ● | N/S | PR | ○ | - |
| | Immediate Failure Attack [121] | | T | 1 | ● | ○ | 2 | ○ | - | | | | | | | | | | | |
| | Latent Failure Attack [121] | | T | 1 | ● | ○ | 3 | ○ | - | | | | | | | | | | | |
| | SABOT [131] | | N | 1 | ● | ○ | 3 | - | - | | | | | | | | | | | |
| | Ladder Logic Exfiltration [53] | Exfiltration over Side-channel | U | 3 | 0 | ○ | 1 | ○ | - | | | | | | | | | | | |
| | Leak Application data [196] | | N | 0 | 0 | ○ | 1 | - | - | | | | | | | | | | | |
| | S7-1200 Download Attack [34] | Program Download | S7 | 0 | 0 | ○ | 2 | ● | - | | | | | | | | | | | |
| | S7-1500 Download Attack [34] | | S7 | 0 | 0 | ○ | 2 | ● | - | | | | | | | | | | | |
| | Denial of Engineering Operations [171] | Adversary-in-the-Middle | E | 1 | 0 | ● | 2 | ○ | - | Optimised Datablocks [159] | Encrypt Sensitive Info. | S7 | 2 | 2 | 1 | ○ | N/S | PR | ○ | - |
| | CaFDI [132] | Brute Force I/O | N | 2 | 0 | ○ | 3 | - | - | | | | | | | | | | | |
| | Time-of-Day Attack [18] | Change Operating Mode | S7 | 1 | 0 | ○ | 2 | ○ | - | PLC redundancy framework [103] | Redundancy of Service | E | 2 | 3 | 2 | ○ | N/S | PR | ○ | - |
| | LogicLocker [70] | Data Encrypted for Impact | M | 0 | 0 | ○ | 2 | ○ | - | Quad-redundant PLC [123] | | X | 2 | 3 | 2 | ○ | N/S | PR | ○ | - |
| | Denial of Decompilation Attack [223] | Denial of Service | S7 | 0 | 0 | ○ | 2 | ○ | - | Digital twin-based simulation [55] | | N | 1 | 3 | 3 | ● | N/S | RE | - | - |
| | Stable Perturbation Attack [165] | ICS Sector Discovery | N | 1 | 0 | ● | 3 | ○ | - | | | | | | | | | | | |
| | Evil PLC Attack [164] | Lateral Tool Transfer | X | 0 | 0 | ○ | 3 | ● | - | | | | | | | | | | | |
| | Targeted Manipulation of FB Operation [88] | Manipulate I/O Image | N | 1 | 0 | ○ | 3 | ● | - | | | | | | | | | | | |
| Firmware | Bricking the device [82] | System Firmware | N | 1 | 0 | ○ | 2 | ○ | - | AttkFinder [39] | Process Vul. Discovery | E | 1 | 1 | 1 | ● | 96 | PR | ○ | - |
| | Compromise System Functions [224] | | SM | 1 | 0 | ○ | 3 | ○ | - | VETPLC [219] | | E | 1 | 1 | 1 | ○ | N/S | PR | ○ | ○ |
| | Decrypting Siemens Simatic firmware [31] | | P | 1 | 0 | ● | 2 | ● | - | Similo [156] | | E | 1 | 2 | 2 | ○ | 100 | RE | ● | - |
| | Firmware leakage [35] | | U | 3 | 0 | ○ | 1 | ○ | - | Logging input simulation [203] | | E | 1 | 2 | 1 | ○ | 99.9 | DE | ○ | - |
| | Firmware modification attack [27] | | U | 3 | 0 | ○ | 3 | ○ | - | CPAC [63] | Auth. Enforcement | X | 2 | 2 | 2 | ○ | N/S | PR | ○ | - |
| | Persistent denial-of-service attack [169] | Modify Program | E | 0 | 0 | ○ | 2 | ○ | - | Firmware verification tool [134] | Code Signing | X | 1 | 1 | 1 | ○ | N/S | PR | ○ | - |
| | Remotely-triggered DoS [169] | | E | 0 | 0 | ○ | 2 | ○ | - | PLCDefender [162] | Attestation | E | 2 | 3 | 1 | ○ | 98 | PR | ● | ○ |
| | Time-based denial-of-service attack [169] | | E | 0 | 0 | ○ | 2 | ○ | - | | | | | | | | | | | |
| | Flash Update [163] | Activate Firmware Update Mode | E | 0 | 0 | ● | 2 | ○ | - | | | | | | | | | | | |
| | HARVEY [75] | Data from Debug Port | MC | 3 | ● | ○ | 3 | ○ | - | | | | | | | | | | | |
| | Shutting down the PLC [82] | Device Restart/Shutdown | N | 1 | 0 | ○ | 2 | ○ | - | | | | | | | | | | | |
| | Rogue Firmware Load [151] | Module Firmware | E | 1 | 0 | ○ | 2 | ● | - | SNIFU [30] | Code Signing | U | 2 | 2 | 1 | ○ | 100 | PR | ○ | - |
| | Leak I/O Process Data [196] | Exfiltration over Side-channel | N | 2 | 0 | ○ | 1 | - | - | ORRIS [157] | Antivirus/Antimalware | X | 2 | 2 | 1 | ● | 85.6 | DE | ○ | C |
| | Blinkware [195] | | X | 3 | 0 | ○ | 1 | ○ | - | | | | | | | | | | | |
| | Analog Emissions attack [114] | | N | 3 | 0 | ○ | 1 | ○ | - | | | | | | | | | | | |
| | PHYCO [76] | | U | 3 | 0 | ○ | 1 | ○ | - | | | | | | | | | | | |
| I/O | Escalated Privilege I/O Command [87] | Adversary-in-the-Middle | E | 2 | 0 | ○ | 2 | ○ | - | Hidden Sensor Measurements [81] | Network Intrusion Detection | X | 1 | 2 | 1 | ○ | N/S | DE | ○ | - |
| | I/O Command attack [87] | | E | 1 | 0 | ○ | 2 | ○ | - | Physics-based attack detection [199] | | E | 1 | 1 | 1 | ○ | N/S | DE | ○ | - |
| | Wireless Control [68] | Spoof Reporting Message | N | 3 | 0 | ○ | 2 | ● | - | Blockchain Monitoring [42] | Encrypt Network Traffic | E | 1 | 3 | 2 | ○ | N/S | PR | ○ | - |
| | Malicious Expanders [68] | | N | 3 | 0 | ○ | 2 | ○ | - | | | | | | | | | | | |
| | False sequence attack [210] | Brute Force I/O | N | 2 | ● | ○ | 3 | - | - | PCAT [174] | Validate Program Inputs | E | 1 | 2 | 1 | ○ | N/S | PR | - | - |
| | Manipulated Variable [12] | Manipulation of Control | X | 2 | 0 | ○ | 2 | ○ | - | Smart I/O Modules [150] | | M | 1 | 2 | 1 | ● | N/S | PR | - | - |
| | OPC UA Supply Chain Attack [94] | Supply Chain Compromise | UA | 0 | 0 | ○ | 2 | ○ | - | PAtt [79] | Attestation | X | 2 | 2 | 1 | ○ | 97 | PR | ○ | - |
| | Backdoor Attack [202] | Unauthorized Command Message | N | 2 | 0 | ● | 2 | - | - | | | | | | | | | | | |
| | Leak Crypto secrets [196] | Exfiltration over Side-channel | N | 1 | 0 | ○ | 1 | - | - | | | | | | | | | | | |
| | WaterLeakage [158] | | S7 | 0 | 0 | ○ | 1 | ○ | - | | | | | | | | | | | |
| Memory | PLC-Blaster [182] | Change Operating Mode | S7 | 0 | 0 | ○ | 2 | ○ | - | Mitigate Malicious Disruption [40] | Redundancy of Service | N | 2 | 1 | 1 | ○ | N/S | DE | - | ○ |
| | Clear PLC memory [205] | Data Destruction | P | 0 | 0 | ● | 2 | ○ | - | CPS Twinning [59] | | M | 1 | 3 | 3 | ● | N/S | RE | ○ | - |
| | ICS-BROCK [221] | Data Encrypted for Impact | U | 3 | 0 | ○ | 2 | ○ | - | Armor PLC [220] | | N | 1 | 3 | 1 | ○ | N/S | DE | - | ○ |
| | Memory Dump [224] | Data from Debug Port | SM | 1 | 0 | ○ | 1 | ○ | - | | | | | | | | | | | |
| | Exfiltrate FB Variables [88] | Exfiltration over ICS Protocol | P | 0 | 0 | ○ | 1 | ● | - | | | | | | | | | | | |
| | Storage Based Covert Channel [88] | Fallback Channels | P | 0 | 0 | ○ | 1 | ● | - | | | | | | | | | | | |
| | DB content manipulation [126] | Modify Parameter | S7 | 0 | 0 | ○ | 2 | ○ | - | | | | | | | | | | | |
| | ROP Attack [26] | Process Injection | M | 0 | 0 | ○ | 2 | ○ | - | | | | | | | | | | | |
| | False Command Injection Attack [14] | Unauthorized Command Message | M | 0 | 0 | ○ | 2 | ○ | ○ | | | | | | | | | | | |
| CPU | Forcing a CPU Stop [163] | Device Restart/Shutdown | E | 0 | 0 | ● | 2 | ○ | - | WeaselBoard [141] | Exploit Protection | BP | 3 | 3 | 3 | ○ | N/S | PR | ● | - |
| | Crash CPU [163] | | E | 0 | 0 | ● | 2 | ○ | - | C2 [130] | | N | 2 | 2 | 1 | ○ | N/S | PR | ○ | - |
| | CPU Stop and Start Attack [31] | | P | 0 | 0 | ● | 2 | ○ | - | | | | | | | | | | | |
| | S7-1200 Start/Stop Attack [34] | Change Operating Mode | S7 | 0 | 0 | ○ | 2 | ● | - | | | | | | | | | | | |
| | S7-1500 Start/Stop Attack [34] | | S7 | 0 | 0 | ○ | 2 | ● | - | | | | | | | | | | | |
| | ASIC Reverse Engineering | Supply Chain Compromise | X | 3 | 0 | ○ | 3 | ○ | - | | | | | | | | | | | |

Table 2: A Summary of PLC Attack and Defense Methods (Control Logic, Firmware, I/O, Memory and CPU Components).
**S7**=S7COMM, **UA**=OPC-UA, **P**=Profinet, **E**=EtherNet/IP, **M**=Modbus TCP, **T**=TriStation, **SM**=SoMachine, **U**=USB Port, **MC**=Memory Card, **X**=Others, **O**=OpenPLC, **C**=CODESYS, **N**=Not Specified, **DE**=Detection, **PR**=Prevention, **RE**=Recovery access to protected actions in the PLC (e.g., the attacker can send start and stop commands to the PLC).

**Attacks that Target the Control Logic.** Two of the most important challenges for modifying the control logic of a PLC are 1) how to infect the PLC without being detected and 2) how to hide the infection from the engineering workstation. To address challenge 1, *Yoo and Ahmed* [216] propose two control logic infection attacks that can bypass network intrusion detection systems. In the first attack, they bypass intrusion detection systems that look for transfers of control logic (compile code) by injecting control code in data blocks (used for the transfer of data such as counters) and then modifying the execution pointer to execute code in data blocks. The second attack uses fragmentation and noise to further obfuscate these control logic transfers to the PLC.

*Kalle et al.* [105] target both challenges (infection and stealthiness). They consider a different problem in code injection, where they assume they have a binary they want to modify. They then develop a decompiler transforming low-level control logic to a high-level instruction list to help them inject the malicious code before recompiling it into a binary that can be uploaded to the vulnerable PLC. They also develop a virtual PLC that interfaces with the engineering workstation (via an AiTM attack); this virtual PLC then sends previously captured network traffic of the original uninfected control logic back to the workstation.

**Attacks that Target the PLC CPU.** As discussed in Appendix E, PLCs generally have three CPU operating modes: "STOP," "RUN," and "PROGRAM." Attacks targeting the PLC CPU focus on disabling the CPU remotely (AL3 or AL2) with STOP commands. One of the earliest examples of these attacks launched against Siemens PLCs by impersonating the workstation is the work of Beresford [31]. While Siemens released cryptographic protections for these connections that would prevent these attacks, more recent work reverse-engineered the cryptographic protocol. *Biham et al.* [34] were able to create a rogue engineering station that could remotely start or stop these newer PLCs by compromising a vulnerable key exchange protocol.

Passwords can protect CPU modes of operation, but a compromised password can still enable remote attacks. Higher-end PLCs protect their CPU modes with physical methods, such as using a physical switch or a physical key. These methods are further discussed in Appendix E.

**Attacks that Target the PLC Firmware.** Firmware modification is one of the most powerful attacks in any platform, as the attacker can control access to the input and output modules of the PLC while remaining undetected. *Basnight et al.* [27] pioneered methods on PLC firmware reverse-engineering and how to develop modified firmware as a proof of concept.

The most cited firmware attack paper in our study is Harvey [75]. The authors extracted firmware images from the update packages of the PLC vendor's website and the PLC memory through the JTAG interface of the PLC processor.

Then they identified the subroutines that allowed them to modify the inputs and outputs for the PLC. To upload the firmware, they rely on vulnerabilities to protect remote firmware update functions (AL0, AL1 or AL2 in Fig. 4) or directly through the JTAG interface (AL3 in Fig. 4).

**Attacks that Target the PLC I/O.** The two most cited papers targeting the inputs and outputs of the PLC focus on how to use input or output physical signals as covert channels to exchange information. For example, *PHYCO* [76] proposes a method where two compromised PLCs can talk to each other, even when a firewall exists between them. One PLC can send an output to increase power generation, and the other PLC can read the increased generation as a bit of information. Adversaries can also use the PLC I/O to exfiltrate data. *Krishnamurthy et al.* [114] illustrate how a PLC sending control commands to a motor can leak information about the status of a chemical plant.

**Attacks that Target the PLC Runtime.** An illustrative example of runtime attacks is provided by *Abbasi et al.* [2], by introducing Pin Control Attacks. This attack involves tampering with the SoC configuration within the PLCs, aiming to disrupt the communication between the PLC runtime and the hardware peripherals. By implementing this attack, the adversary severs the connection between the runtime software and the physical world, allowing them to manipulate the control data within the actual physical process.

Runtime attacks can also target the availability of the system; for example, *Gjendemsjø et al.* [82] proposed an XML Bomb Attack that causes the PLC runtime to crash by modifying an XML file.

**Attacks that Target the PLC Operating System.** There are not a lot of papers focusing on attacking the OS of PLCs. In their comprehensive study of the attack surface of a PLC, *Abbasi et al.* [4] touch upon an intriguing aspect of PLC security: the vulnerability of the Siemens Adonis Real-Time Operating System. Although their primary focus lies on the bootloader security of Siemens PLCs, they also explore security weaknesses inherent in the Adonis RTOS.

**Attacks that Target the PLC Memory.** Some network or control logic attacks target the memory of the PLC as part of their infection chain by modifying memory blocks or by getting memory dumps. The most cited efforts focused on developing worms stored in memory [158, 182].

So far, we have introduced the most cited examples of attacks per target, but we have not illustrated representative defenses. Because of limited space, we refer the reader to a discussion of defenses in Appendix G.

## 6 Research Gaps

We now focus on identifying relevant insights, gaps, and recommendations for future work by analyzing the data from Tables 1 and 2.

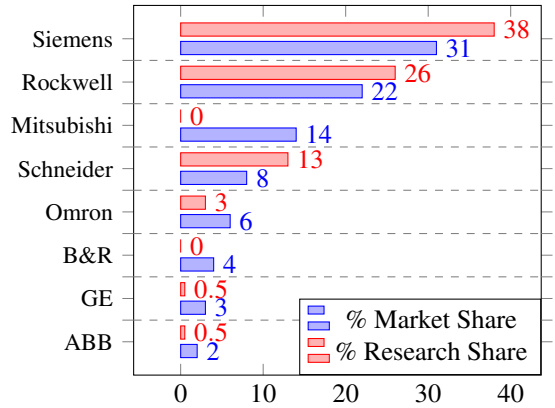

Figure 5: Comparison of Research Share vs Market Share.

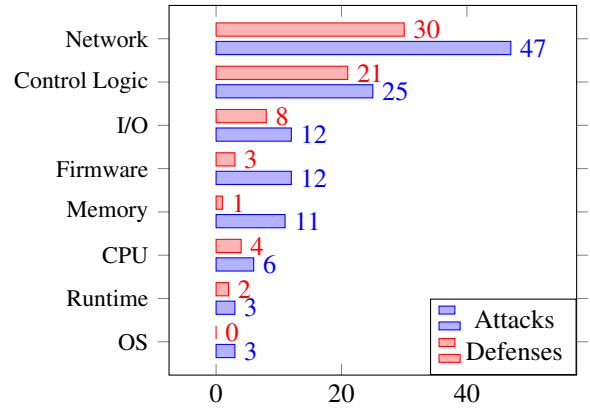

Figure 7: Defense vs Attack Methods per Target Component.

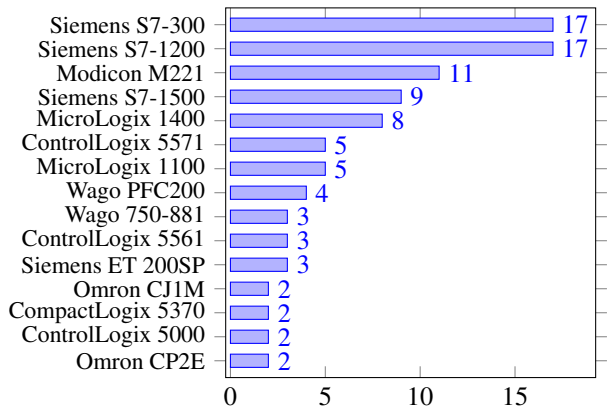

Figure 6: The 15 Most Common PLC Models.

**Most of the Attacks Require Zero Environment Knowledge.** We found that 82% (97/119) of the attacks that may cause either *limited* (①) (36% (35/97)), *substantial* (②) (56% (54/97)), or *severe* (③) (8% (8/97)) damage require *zero knowledge* (○) of the environment in which the PLC lives, following the description presented in Sec. 2.1 and the criteria discussed in Sec. 4.3. This may imply that adversaries can potentially launch non-trivial attacks against power grids, chemical plants, water treatment plants, etc. by targeting PLCs without having to invest time performing reconnaissance or learning the specifics of actuators, sensors, etc. Moreover, we were able to identify that only a few defense methods, e.g., *Formby et al.* [69] and *Bellettini et al.* [28], make use of such important environmental information. We therefore recommend that future defenses make use of environment knowledge in their strategy to increase their effectiveness.

**The Security of Important PLC Brands Has Not Been Explored.** We found that the security of some important PLC platforms widely used in practice has been ignored. As shown in Fig. 5, the *market* share percentage of important PLC manufacturers such as Mitsubishi, Omron, ABB, and GE is considerably higher than their *research* share, e.g., the number of papers explicitly using them for attack/defense evaluation purposes. For example, even though Mitsubishi PLCs account for 14% of the global market, they contribute to 0% of the research share in our review (they only appear once in our Grey literature survey [41]).

This means that there are potentially thousands of PLCs deployed in the world whose security has not been explored or may not have been adequately understood. We therefore recommend future lines of work specifically addressing these devices, exploring the effectiveness of existing attacks and defenses as discussed in this work, as well as the introduction of newer security methods tailored for them.

**Lack of Defenses at the Recovery Stage.** We found that most defense methods are designed for the Prevention (PR) and Detection (DE) stages discussed in Sec. 4.7, 47% (33/70) and 47% (33/70) out of 70 respectively, whereas the Recovery (RE) stage accounts for only 6% (4/70). This means that in the event of a successful attack there are limited options to recover and bring the PLC back to operation.

**Most Attacks and Defenses are Evaluated on a Small Subset of PLCs.** Our results show that the top 5 most common PLC models in our literature review (as shown in Fig. 6) account for 40% of all studies. This may result in a narrow understanding of PLC security that excludes the rest of the PLC models in the market. Additionally, we found that 80% (95/119) of attack methods and 81% (57/70) of defense methods were evaluated using a single HardPLC model (○). This means that most attack and defense methods are shown to work with a single HardPLC model, making it unclear whether or not the defense method can be generalized to other PLC models and manufacturers. Therefore, we recommend that future research include an evaluation of multiple PLCs.

**Important Tactics have Little to No Research.** Our ICS$^2$ Matrix includes 14 Tactics. However, as Fig. 8 shows, most of the attack methods focus on Impair Process Control, Inhibit Response Function, Exfiltration, Collection, and Persistence. On the other hand, important Tactics like Command and Control, Lateral Movement, Evasion, Credential Access, and Initial Access have not been investigated.

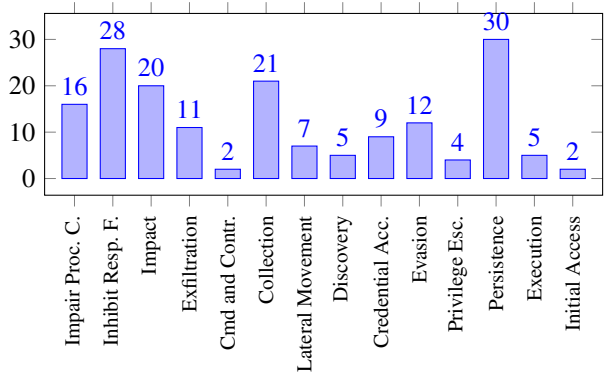

Figure 8: Attack Methods according to their Tactic.

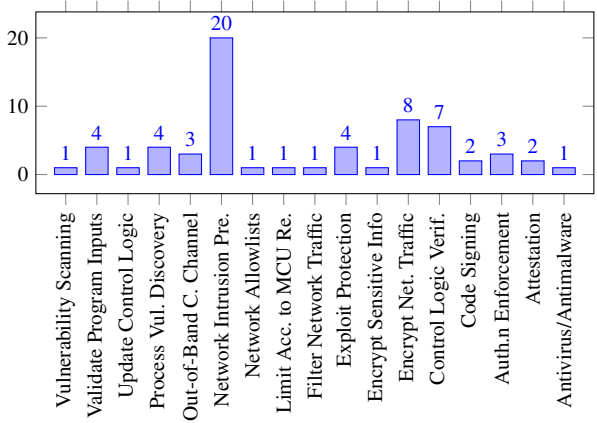

Figure 9: Defense Methods per Mitigation Category.

One reason that might explain why these tactics have not been explored in previous research is that they are too specific or custom-fit. For example, the Command and Control tactic includes techniques such as *Commonly Used Port* and *Standard Application Layer Protocol*, which depend on specific network protocols and ports. This specific focus on network ports might limit the scope of research that can be carried out for this tactic.

**Most Mitigation Strategies have Little to No Research.** We found defense methods that match 17 Mitigation categories. However, as shown in Fig.9, the majority of these defenses fall into the Network Intrusion Prevention, Encrypt Network Traffic and Control Logic Verification categories. The remaining 14 Mitigation categories have only 4 or fewer defenses. These include important categories like Antivirus/Antimalware, Vulnerability Scanning, and Attestation, which are especially important given the rise of ICS malware such as the ones described in Appendix F.

**Weaknesses of State-of-the-Art Defenses** Based on the results shown in Tables 1 and 2, we identified the following three major weaknesses when it comes to defending PLCs: **1)** No ransomware detection. Even though researchers have shown that ransomware attacks against PLCs are possible [70, 221] and there are multiple documented ransomware attacks against ICS, there is no available research focused on how to detect and stop PLC ransomware. To address this research gap, future research should introduce new defense methods that take advantage of state-of-the-art malware and ransomware detection techniques such as sandbox detection [200] and static and dynamic analysis [148]. **2)** No web-based malware detection. Research shows that attackers target PLCs' web interfaces [122, 166] and that it is possible to compromise PLCs via their web applications [152]. However, there is no available research focused on how to detect and stop malware targeting PLCs' web interfaces. **3)** No Exfiltration over Covert Channel Detection. This is one of the new techniques that we introduced in our $ICS^2$ Matrix (Fig. 13), which includes methods such as *PHYCO* [76]. Introducing this technique lays the groundwork for identifying the need for mitigations against such techniques. Currently, there is no known mitigation for this type of attack. Indeed, in 2014 *Garcia et al.* wrote: "There has been no detection solution capable of identifying such hidden communications in the physical power system," [76] which still holds true.

## 7 Discussion

We now discuss research challenges that have not attracted enough attention and may become relevant as PLCs evolve.

**Reproducible Research.** Based on our analysis and the results presented in Tables 1 and 2, few defense and attack methods provide publicly available research artifacts. During our literature review, we searched for research artifacts for each paper. We searched the paper itself on Google and the author's website. Using this method, we were able to find the source code for 19 papers (only 16%). This limits the reproducibility of attacks and defenses. In an attempt to find the artifacts for papers without openly available code, we contacted 91 authors via email requesting their research artifacts, and we received 16 responses (17.6%). Ultimately only 3 shared a research artifact. The other 13 did not share the source code for the following reasons: **1)** The project was completed long ago or the first author moved on to a different institution (30%). **2)** There were funding or distribution restrictions (25%). **3)** The authors were working on it and will publish it later (15%). **4)** There were no plans to release it to the public (30%). 16% is a low number of papers with artifacts, and this does not even consider if the source code has good documentation or if these public resources are easy to run. Therefore, we encourage researchers to release their PLC security artifacts so that research can be replicated and built upon and to leverage our PLC security artifacts repository discussed in Sec. 1 to disseminate their artifacts. We acknowledge that given the criticality of PLCs it is not always possible to release artifacts at the time of publication if at all. For example, the US' Cybersecurity & Infrastructure Security Agency (CISA) recommends researchers "provide us a reasonable amount of time to resolve the issue before you disclose it publicly [201]."

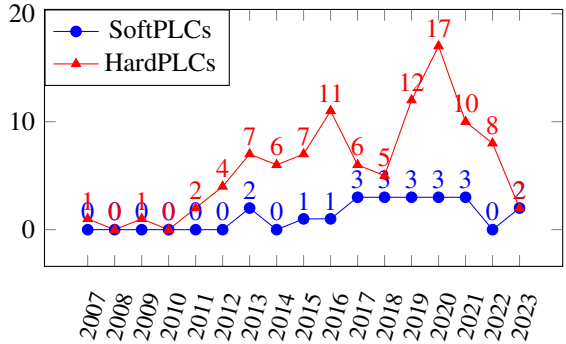

Figure 10: Papers evaluated using SoftPLCs vs HardPLCs.

A second challenge that compounds the problem of lack of reproducible research is the absence of standardized performance metrics. Based on the results in Tables 1 and 2, few defense methods reported detailed PLC overhead or effectiveness metrics as described in Sec. 4.7. This makes comparing and building upon previous research almost impossible. To overcome this challenge we recommend that future defense methods report detailed quantifiable metrics. First, papers that propose methods such as Intrusion Detection Systems (IDS) should report a complete and detailed confusion matrix. The work by *Salehi et al.* [162] provides an excellent example. Second, future research on defense methods should report detailed and quantifiable benchmark metrics, e.g., memory overhead. Existing benchmarks such as *BenchmarkIoT* [13] can be customized for PLCs to overcome this challenge.

**Transition from HardPLCs to SoftPLCs.** As discussed in the introduction, PLCs are going through a paradigm shift with support for new protocols and new functionalities. Support for SoftPLCs appears to be increasing. While SoftPLCs currently have a minimal market share, large manufacturers like Siemens and Rockwell are preparing hardware-agnostic products [180]. This paradigm shift has also reached the scientific community. As shown in Fig. 10, PLC security research has diverged into two different strands: evaluations that use HardPLCs and evaluations that use SoftPLCs.

Considering this trend, future research should be aimed at addressing the following challenges: **1)** Developing transitional defense methods that secure both HardPLCs and SoftPLCs. For example, developing *bump-in-the-wire* solutions that are compatible with HardPLCs and SoftPLCs. This may involve evaluating research on both HardPLCs and SoftPLCs. **2)** Investigating defense mechanisms available for SoftPLCs that were previously unavailable for HardPLCs. For example, HardPLCs' proprietary nature limited their access control and encryption capabilities by incorporating weak encryption protocols. SoftPLCs' less opaque architecture provides the opportunity to solve this problem. **3)** Investigating both attack and defense methods that are possible only with SoftPLCs. For example, CODESYS' SoftPLCs can be connected to the cloud [45], which opens new ways to collect data that can be used to train machine learning-based Intrusion Detection Systems.

## 8 Related Work

**SoKs on PLCs.** To the best of our knowledge, the work by *Sun et al.* [188] is the only SoK focused on attacks and defenses on the control logic of PLCs. However, our work is different in two ways. First, *Sun et al.* focuses on PLCs' control logic, while we take a more comprehensive approach that includes 8 additional PLC components. Second, our SoK includes papers with practical evaluations, while *Sun et al.* focus on formal or theoretical research.

**PLC Honeypots.** While previous work includes approaches for PLC honeypots [72, 122], their focus is intelligence gathering rather than attacking PLCs.

**Fuzzing and Binary Reverse Engineering.** There is also noticeable research on fuzzing and reverse engineering for ICS and PLCs in the literature [66, 107, 145, 194, 204]. Our focus, however, is not on software analysis or patching but on outlining specific attacks.

**Embedded Controllers.** While they have received most of the attention in the security literature, PLCs are not the only embedded equipment in ICS. Remote Terminal Units (RTUs) and Intelligent Electronic Devices (IEDs) are even more prevalent in energy transmission systems and substation automation [149, 161, 187]. We need further research into the security of these other embedded controllers.

## 9 Conclusion and Future Work

In this paper, we provide a systematic review of currently-known attacks and defenses for PLCs. We also pointed out research gaps that should be tackled in the future so that the security of PLCs can be better understood, thus helping avoid future attacks against ICS and PLCs. We hope this systematization is useful to newcomers to the field as well as experienced PLC researchers looking to contextualize their work. As a part of Future Work, we are developing an experimental testbed as discussed in Sec. 2.1. This way, different attack and defense strategies can be replicated to provide experimental evidence on their effectiveness, deployability, and robustness, thus ultimately complementing the results provided in this work.

## Acknowledgments

This research is partially supported by NSF awards CNS-1929410, CNS-1931573, CNS-2150351, and CNS-2131263.

[31] BERESFORD, D. Exploiting siemens simatic s7 plcs. *Black Hat USA 16*, 2 (2011), 723–733. https://media.blackhat.com/bh-us-11/Beresford/BH_US11_Beresford_S7_PLCs_WP.pdf. Accessed: 27-09-2023.

[32] BHATIA, S., KUSH, N. S., DJAMALUDIN, C., AKANDE, A. J., AND FOO, E. Practical modbus flooding attack and detection. In *Proceedings of the Twelfth Australasian Information Security Conference (AISC 2014)[Conferences in Research and Practice in Information Technology, Volume 149]* (2014), Australian Computer Society, pp. 57–65.

[33] BIALLAS, S., BRAUER, J., AND KOWALEWSKI, S. Arcade. plc: A verification platform for programmable logic controllers. In *2012 Proceedings of the 27th IEEE/ACM International Conference on Automated Software Engineering* (2012), IEEE, pp. 338–341.

[34] BIHAM, E., BITAN, S., CARMEL, A., DANKNER, A., MALIN, U., AND WOOL, A. Rogue7: Rogue engineering-station attacks on s7 simatic plcs. *Black Hat USA* (2019). https://i.blackhat.com/USA-19/Thursday/us-19-Bitan-Rogue7-Rogue-Engineering-Station-Attacks-On-S7-Simatic-PLCs-wp.pdf. Accessed: 27-09-2023.

[35] BITAN, S., AND DANKNER, A. sOfT7: Revealing the Secrets of the Siemens S7 PLCs. *Black Hat USA* (2022), 36. https://i.blackhat.com/USA-22/Wednesday/US-22-Bitan-Revealing-S7-PLCs.pdf. Accessed: 27-09-2023.

[36] BOLTON, W. *Programmable logic controllers*. Newnes, 2015.

[37] BONNEY, G., HÖFKEN, H., PAFFEN, B., AND SCHUBA, M. Ics/scada security analysis of a beckhoff cx5020 plc. In *2015 International Conference on Information Systems Security and Privacy (ICISSP)* (2015), IEEE, pp. 1–6.

[38] BRUSSO, B. C. 50 years of industrial automation [history]. *IEEE Industry Applications Magazine 24*, 4 (2018), 8–11.

[39] CASTELLANOS, J. H., OCHOA, M., CARDENAS, A. A., ARDEN, O., AND ZHOU, J. Attkfinder: Discovering attack vectors in plc programs using information flow analysis. In *24th International Symposium on Research in Attacks, Intrusions and Defenses (RAID 2021)* (San Sebastian, Spain, 2021), RAID '21, Association for Computing Machinery, p. 235–250.

[40] CHATTERJEE, U., SANTIKELLUR, P., SADHUKHAN, R., GOVINDAN, V., MUKHOPADHYAY, D., AND CHAKRABORTY, R. S. United we stand: A threshold signature scheme for identifying outliers in plcs. In *2019 56th ACM/IEEE Design Automation Conference (DAC)* (2019), IEEE, pp. 1–2.

[41] CHENG, M., AND YANG, S. Taking Apart and Taking Over ICS & SCADA Ecosystems: A Case Study of Mitsubishi Electric. *DEF CON 29* (2021). https://media.defcon.org/DEF%20CON%2029/DEF%20CON%2029%20presentations/Mars%20Cheng%20Selmon%20Yang%20-%20Taking%20Apart%20and%20Taking%20Over%20ICS%20%26%20SCADA%20Ecosystems%20-%20%0A%20Case%20Study%20of%20Mitsubishi%20Electric.pdf. Accessed: 03-10-2023.

[42] CHOI, M. K., YEUN, C. Y., AND SEONG, P. H. A novel monitoring system for the data integrity of reactor protection system using blockchain technology. *IEEE Access 8* (2020), 118732–118740.

[43] CHOI, T., BAI, G., KO, R. K., DONG, N., ZHANG, W., AND WANG, S. An analytics framework for heuristic inference attacks against industrial control systems. In *2020 IEEE 19th International Conference on Trust, Security and Privacy in Computing and Communications (TrustCom)* (2020), IEEE, pp. 827–835.

[44] Cloud Connectivity – IoT PLC: Controllers with MQTT. https://www.wago.com/us/open-automation/cloud-connectivity/iot-plc-controller-with-mqtt. Accessed: 27-09-2023.

[45] Codesys automation server | codesys store international. https://store.codesys.com/en/codesys-automation-server.html. Accessed: 08-11-2023.

[46] CODESYS Inside. https://www.codesys.com/the-system/codesys-inside.html. Accessed: 27-09-2023.

[47] CODESYS Runtime. https://www.codesys.com/products/codesys-runtime.html. Accessed: 27-09-2023.

[48] CONN, V. S., VALENTINE, J. C., COOPER, H. M., AND RANTZ, M. J. Grey literature in meta-analyses. *Nursing research 52*, 4 (2003), 256–261.

[49] ControlLogix System User Manual. https://literature.rockwellautomation.com/idc/groups/literature/documents/um/1756-um001_-en-p.pdf. Accessed: 02-10-2023.

[50] COOK, M. M., MARNERIDES, A. K., AND PEZAROS, D. Plcprint: Fingerprinting memory attacks in programmable logic controllers. *IEEE Transactions on Information Forensics and Security 18* (2023), 3376–3387.

[51] Cyber-Physical Systems Workshop. https://web.archive.org/web/20080517071555/http://varma.ece.cmu.edu/cps/, 2006. Accessed: 27-09-2023.

[52] CYBERSECURITY, AND (CISA), I. S. A. Apt cyber tools targeting ics/scada devices. https://www.cisa.gov/uscert/ncas/alerts/aa22-103a, 2022. Accessed: 02-10-2023.

[53] DAVID, A., AND GEORGE, L. Exfiltrating reconnaissance data from air-gapped ics/scada netowrks. *Black Hat Europe* (2017). https://www.blackhat.com/docs/eu-17/materials/eu-17-Atch-Exfiltrating-Reconnaissance-Data-From-Air-Gapped-Ics-Scada-Networks.pdf. Accessed: 02-10-2023.

[54] DIETRICH, D., NEUMANN, P., AND SCHWEINZER, H. F. Fieldbus technology: Systems integration, networking, and engineering. In *Proceedings of the Fieldbus Conference FeT'99 in Magdeburg, Federal Republic of Germany* (1999), Springer.

[55] DIETZ, M., VIELBERTH, M., AND PERNUL, G. Integrating digital twin security simulations in the security operations center. In *Proceedings of the 15th International Conference on Availability, Reliability and Security* (New York, NY, USA, 2020), ARES '20, Association for Computing Machinery.

[56] DRAGOS, A. Trisis malware: Analysis of safety system targeted malware. https://www.dragos.com/resource/trisis-analyzing-safety-system-targeting-malware/, 2017. Accessed: 27-09-2023.

[57] DRAGOS, I. Chernovite's pipedream malware targeting industrial control systems (ics). https://www.dragos.com/blog/industry-news/chernovite-pipedream-malware-targeting-industrial-control-systems/, 2022. Accessed: 27-09-2023.

[58] DUNLAP, S., BUTTS, J., LOPEZ, J., RICE, M., AND MULLINS, B. Using timing-based side channels for anomaly detection in industrial control systems. *International Journal of Critical Infrastructure Protection 15* (2016), 12–26.

[59] ECKHART, M., AND EKELHART, A. Towards security-aware virtual environments for digital twins. In *Proceedings of the 4th ACM Workshop on Cyber-Physical System Security* (New York, NY, USA, 2018), CPSS '18, Association for Computing Machinery, p. 61–72.

[60] Encrypted block protection for fbs and fcs from step 7 v5.5. https://support.industry.siemens.com/cs/document/45632073/how-do-you-install-the-improved-block-protection-for-fbs-and-fcs-in-step-7-v5-5-onwards-?dti=0&lc=en-AO. Accessed: 02-10-2023.

[61] ERBA, A., TAORMINA, R., GALELLI, S., POGLIANI, M., CARMINATI, M., ZANERO, S., AND TIPPENHAUER, N. O. Constrained concealment attacks against reconstruction-based anomaly detectors in
15

# Appendices

## Appendix A  Scientific and Grey Literature Resources

The seven digital libraries queried during our scientific literature review are as follows: ACM Digital Library[7], arXiv[8], dblp[9], Google Scholar[10], IEEExplore[11], USENIX Papers Search[12], and NDSS Symposium Search[13].

The three main sources we queried during our grey literature review are as follows: Digital Bond Archives[14], InfoconDB[15], and Google[16].

**Search keywords.**

```
"plc", "programmable logic controller", "scada", "cps",
"cyber-physical system", "cyber physical system", "iiot",
"industrial internet of things", "industry 4.0",
"industrial control system", "ics", "embedded system",
"attack", "threat", "vulnerability", "defense"
```

## Appendix B  MITRE Technique and Sub-Technique Model Example

As an example of how our ICS$^2$ Matrix can be used to categorize ICS-specific attacks proposed by researchers, we will have a look at the work by *Krishnamurthy et al.* [114] They proposed an attack strategy where a malware-infected PLC can exploit the acoustic emissions generated by a motor controlling a valve in a feedback control loop within an ICS. This covert acoustic channel enables the malware to secretly transmit sensitive information, such as proprietary controller parameters and system passwords, to a remote receiver. This attack does not disrupt the stability, performance, or signal characteristics of the closed-loop process, making it stealthy and effective at exfiltrating data from the compromised ICS. We categorized this attack under the *Exfiltration* tactic but could not find a fitting technique. Thus, we propose the addition of the technique *Exfiltration over a Covert Channel*. This way of data exfiltration bypasses the techniques *Filter network Traffic* and *Data Loss Prevention*, as covert channels utilize unintended means of communication by definition. To mitigate covert channels requiring the propagation of wireless signals, *Minimize Wireless Signal Propagation* represents a defense, but due to the different forms that covert channels can take, complete mitigation proves difficult, which puts this technique in the category *Mitigation Limited or Not Effective*. Table 3 shows the new technique definition motivated by the example above. This technique definition follows MITRE's official guidelines [137].

## Appendix C  PLCs' Larger Context and Underlying Architecture

In this Appendix, we expand on the environment in which PLCs operate. Specifically, we use the Purdue model [208] to describe the network architecture of an ICS process and where PLCs fit in.

As we discuss in Sec. 1, PLCs control physical machines or actuators, such as pumps. However, these actuators do not exist in a vacuum. They are one component of many that work together to complete a larger industrial process, for example, a water treatment process. Figure 11 depicts a sample of a water treatment process ICS using the Purdue Model.

## Appendix D  PLC Memory Blocks

PLCs have specialized memory blocks that store different types of data. These blocks are made available to applications that perform operations like program upload and download [36] and can be classified as:

---

[7] https://dl.acm.org/
[8] https://arxiv.org/
[9] https://dblp.org/
[10] https://scholar.google.com/
[11] https://ieeexplore.ieee.org/Xplore/home.jsp
[12] https://www.usenix.org/publications/proceedings/
[13] https://www.ndss-symposium.org/
[14] https://dale-peterson.com/digital-bond-archives/
[15] https://infocondb.org/
[16] https://www.google.com/

| Data Item | |
|---|---|
| Name | Exfiltration over Covert Channel |
| Tactic | Exfiltration |
| Data Sources | |
| Description | Adversaries may attempt to exfiltrate data via a covert channel such as analog emissions of physical instrumentation. For example, actuators, sensors, and mechanical structures. These analog emissions can be acoustic or electromagnetic. In some circumstances, the adversary needs to have physical access to the ICS to measure the analog signal using an antenna, for example. The physical medium or device could be used as the final exfiltration point or to hop between otherwise disconnected systems. |
| Asset | Field Controller/RTU/PLC/IED |
| Defense Bypassed | Data Loss Prevention, Filter Network Traffic |
| Contributor | Efrén López Morales |
| Procedure Example | Krishnamurthy, Prashanth, et al. "Process-aware covert channels using physical instrumentation in cyber-physical systems." [114] IEEE Transactions on Information Forensics and Security 13.11 (2018): 2761-2771. |
| Mitigation | Minimize Wireless Signal Propagation, Mitigation Limited or Not Effective |
| Detection | |

Table 3: Proposed technique definition of Exfiltration over Covert Channel according to MITRE ATT&CK: Design and Philosophy [137]

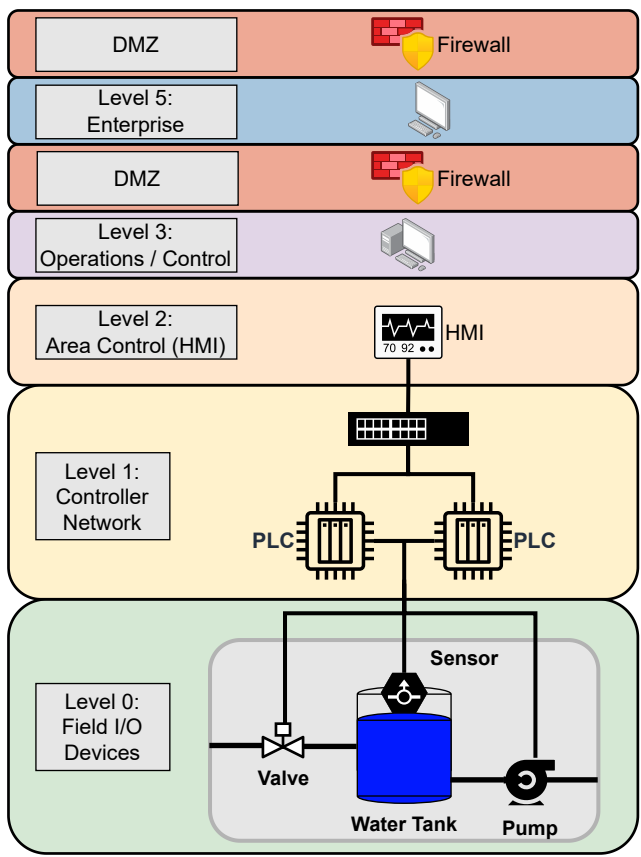

Figure 11: Example of an ICS process with PLCs in Level 1: Controller Network as defined by the Purdue Model. Based on [80, 186].

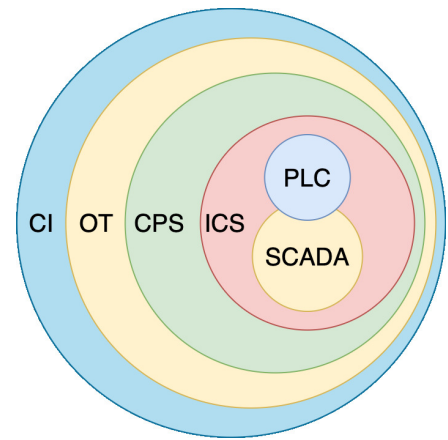

Figure 12: Relationship between Critical Infrastructure (CI), Operational Technology (OT), Cyber-Physical Systems (CPS), Industrial Control Systems (ICS), Supervisory Control and Data Acquisition (SCADA), and PLCs. Based on [139]

**Data Blocks (DB).** These blocks are used to store data that will later be used by a program. Different data types can be stored (e.g., Boolean, byte, integer) [7].

**System Data Blocks (SDB).** These blocks contain PLC configuration information [7,65], listing PLC model, firmware version, IP address, and information about the attached add-ons (e.g., communication processors, frequency converters). SDBs are automatically created and compiled by the PLC and cannot be modified by the user.

**Organization Blocks (OB).** These blocks are the interfaces between the operating system and the user program [6, 120]. OBs execute when events occur (e.g., at CPU startup, clocked executions, errors, hardware interrupts). Several other OBs serve specific roles.

**Function Blocks (FB).** Finally, FBs hold standard executable code blocks, which are written in any of the IEC 61131 standard programming languages [7, 65].

## Appendix E  PLC Built-in Security Features

**Access Control.** Most PLCs include built-in access control features. For example, Siemens and Allen-Bradley PLCs can be configured with a password that restricts changes to the CPU configuration [23, 97]. However, these built-in features have been shown to be ineffective [205].

**Encryption.** Some PLCs also include encryption features for memory blocks, for example, Siemens allows the encryption of specific memory blocks [60]. Another encryption functionality involves the use of hashing to detect changes. Some Rockwell and Siemens PLCs support hashing [193]. Additionally, some of the PLC Industrial Ethernet protocols incorporate built-in encryption. However, like the access control features, they have been shown to be vulnerable [120].

**Operational Modes.** PLCs have different operational modes intended for different scenarios. For example, to modify the control logic program of a production PLC, it is necessary to change or *switch* the PLC operational mode from *Run* to *Stop* and then to *Program* [96]. In practice, there exist two ways to switch the operational mode of a PLC: either to use the PLC management software, e.g., SIMATIC STEP 7 for Siemens PLCs, or to manually switch the operational mode using a physical *key* inserted in the PLC chassis, which overrides the software option just described.

Overall, the most common operational modes are:

- *Run Mode.* This mode is used to execute the control logic program of the PLC. Input from sensor devices is monitored, and output is sent to actuator devices. Ideally, a production PLC should always be set to *Run* [49, 96].
- *Stop Mode.* In this mode, the PLC stops reading inputs and stops the control logic program execution. Typically, a PLC must be switched to *Stop* before it can be re-configured by means of the *Program* mode [91, 96].
- *Program Mode.* In this mode, the PLC control logic program is loaded, modified, or deleted. While in this mode, all outputs from the PLC are stopped [49, 96].

**Disabling Unused Protocols.** Some PLCs have the option to disable network services to reduce unnecessary attack surfaces. For example, Siemens PLCs allow for the integrated web server to be disabled [64], preventing the exploitation of web server-based vulnerabilities [31, 136].

**Monitoring HMI Data.** HMIs are useful to visualize trends about the performance of PLCs, as they can plot the PLC Scan Cycle, the uptime, and shut down and restarts, which can be helpful to detect an ongoing attack [193].

| Access Level | Vector |
|---|---|
| AL3: Physical | • JTAG Port<br>• USB Port<br>• Key Switch<br>• Backplane<br>• Memory Card |
| AL2: Fieldbus | • PROFIBUS<br>• DeviceNet<br>• RS-232 Serial<br>• EtherNet/IP |
| AL1: LAN | • GOOSE<br>• Workstation<br>• SoMachine (Schneider)<br>• TriStation (Schneider)<br>• LS Proprietary (LS Electric) |
| AL0: Internet | • Modbus TCP<br>• DNP3<br>• IEC 104<br>• OPC UA<br>• EtherNet/IP<br>• S7comm<br>• Webserver (HTTP/HTTPS)<br>• SNMP |

Table 4: Attack Vectors per Access Level.

## Appendix F  Real-World PLC Attacks

We summarize real-world attacks that showcase the pressing need to improve the security of PLCs.

**Stuxnet.** Stuxnet is the first-ever documented malware for PLCs. At the time, it set itself apart from previous malware by showing a high level of sophistication, a deep understanding of industrial processes, and the use of four *zero-day* exploits [65]. After compromising a computer with the software for programming PLCs, the malware uploaded its malicious control program to the target PLCs. In particular, the attack targeted Siemens 315 and 417 and made them damage centrifuges while reporting that everything was normal [118].

**Triton.** This malware, also known as TRISIS and HatMan [56, 99] was identified in 2017 after a petrochemical facility in Saudi Arabia was shut down. After compromising an engineering workstation, Triton was able to launch a dropper (trilog.exe) to deliver backdoor files to Safety Instrumented System (SIS) PLC. The first backdoor file was a 0-day exploit that allowed the attackers to inject the second file into the PLC's memory. With a program in the memory of the PLC, the attackers could have control of the device. The attackers were unable to take full control of the system because an error in the PLC caused a system shutdown.

**Pipedream Toolkit.** At the time of writing, Pipedream (also known as Incontroller) is the latest documented malware that specifically targets PLCs [57]. This is not a single purpose malware but a modular framework that includes multiple exploits that target different PLCs. Once the attackers compromise a computer in the control network, Pipedream can be used to scan and compromise Scheider Electric PLCs,

OMROM Sysmac NEX PLCs, and Open Platform Communications Unified Architecture (OPC UA) servers. Pipedream is believed to have been developed by a nation state [52].

**Crashoverride.** Also known as Industroyer [161, 179], Crashoverride is believed to have caused the power outage in Ukraine's capital in December of 2016 [93]. A sophisticated malware designed to disrupt ICS networks used in electrical substations, it targeted the Open Platform Communications Data Access protocol or OPC-DA for short, which defines how data can be transferred to and from PLCs.

## Appendix G  Overview of Defenses

**Defenses Focusing on Network Inspection.** Since most of the attacks in the literature focus on network exploits, it is natural to expect several defenses in the network as well. Some of the first and most popular network intrusion detection systems proposed for industrial networks model the highly-periodic network traffic as a deterministic finite automaton (DFA) and propose to raise alerts when the network traffic does not follow the learned DFA [110]. Other approaches for protecting the networks of PLCs include adding cryptographic protections and machine learning for anomaly detection and prevention [20].

**Control Logic Defenses.** Ensuring that the control program running on the PLC is verified and correct is an old area of research, as it is not necessarily related to security but focuses on providing safety guarantees for the operation of a process [33]. This area has received renewed attention in the security community. An example of this line of research is illustrated by *McLaughlin et al.* [133], which proposed a Trusted Safety Verifier approach that verifies the control logic code through a bump-in-the-wire before it is executed by the PLC.

**CPU Defenses.** One way to detect attacks against the CPU module is to monitor all communications being exchanged between the modules of a PLC. WeaselBoard [141] is a backplane analysis system that forwards all inter-module traffic to an analysis system to detect a variety of attacks. This approach is more general than just detecting CPU attacks, but because it captures information directly from the CPU, we think it fits better in this category. Other papers on CPU defenses focus on resiliency. For example, *Luo et al.* [123] proposed Quad-Redundand PLC, a redundancy framework to provide resiliency after one CPU is attacked, aiming for a second CPU to keep the system running.

**Firmware Modification Defenses.** Bump-in-the-wire defenses are general mechanisms where an extra device is placed between the sender and the receiver. This device checks that the data or code sent to the receiver is correct. Some of the more popular firmware defenses are based on this architecture, where the proxy device is used to inspect firmware updates before they are installed in the PLC [30, 134].

**I/O Defenses.** The two most-cited defenses against the inputs and outputs of the PLC include a defense strategy [199] and a prevention mechanism [130]. The work of Urbina et al. [199] focuses on using physical models of the process under control to detect inconsistencies between the inputs and outputs. On the other hand, $C^2$ [130] focuses on preventing malicious outputs from a compromised PLC from reaching actuators. $C^2$ proposes a way to express and enforce security policies, and it also denies actions that violate the policy.

**Runtime Defenses.** ECFI [3] is a control-flow integrity monitoring runtime protection for real-time PLCs. Fundamentally, ECFI achieves this by segregating control-flow verification of the PLC runtime software from control-flow tracing, ensuring the preservation of real-time requirements crucial for PLC operations. Distinguishing itself from other control-flow monitoring systems, ECFI refrains from terminating the runtime process; instead, it promptly notifies the operator of any detected control-flow violations. This approach strikes a balance between protection and timeliness, allowing for effective threat detection without disrupting the overall system functionality. On the other hand, Ghostbuster [1] serves as a defense mechanism designed to identify Pin Control Attacks targeting PLC runtimes [2]. Implemented as a kernel driver, Ghostbuster operates in two distinct modes to ensure comprehensive protection. In Kernel mode, it actively detects alterations made to the SoC debug registers, enabling the detection of elusive "Ghost in the PLCs" I/O intercepts. On the other hand, to thwart user-mode pin control attacks, Ghostbuster employs a distinct strategy. It diligently monitors the SoC configuration of the Pin Control Subsystem, constantly searching for configuration violations that could sever the connection between the PLC runtime and the physical world. By adopting these proactive measures, Ghostbuster effectively fortifies PLC systems against Pin Control Attacks and reinforces the security of the runtime environment.

**OS Defenses.** Another IT protection being considered is malware detection at the OS level [157], which requires new hardware (Power Debug PRO) and could be circumvented using data hooking or gaining kernel privilege level.

**Memory Defenses.** Applying protections like ASLR is now common in most IT infrastructure, but similar approaches are being explored for PLCs. For example, *Robles-Durazno et al.* [159] proposed Optimized Datablocks, an approach in which the allocation of the datablock data is randomized such that the attacker does not know its exact location, making attacks more difficult as with other moving target defenses.

## Appendix H  Excerpt of the ICS$^2$ Matrix

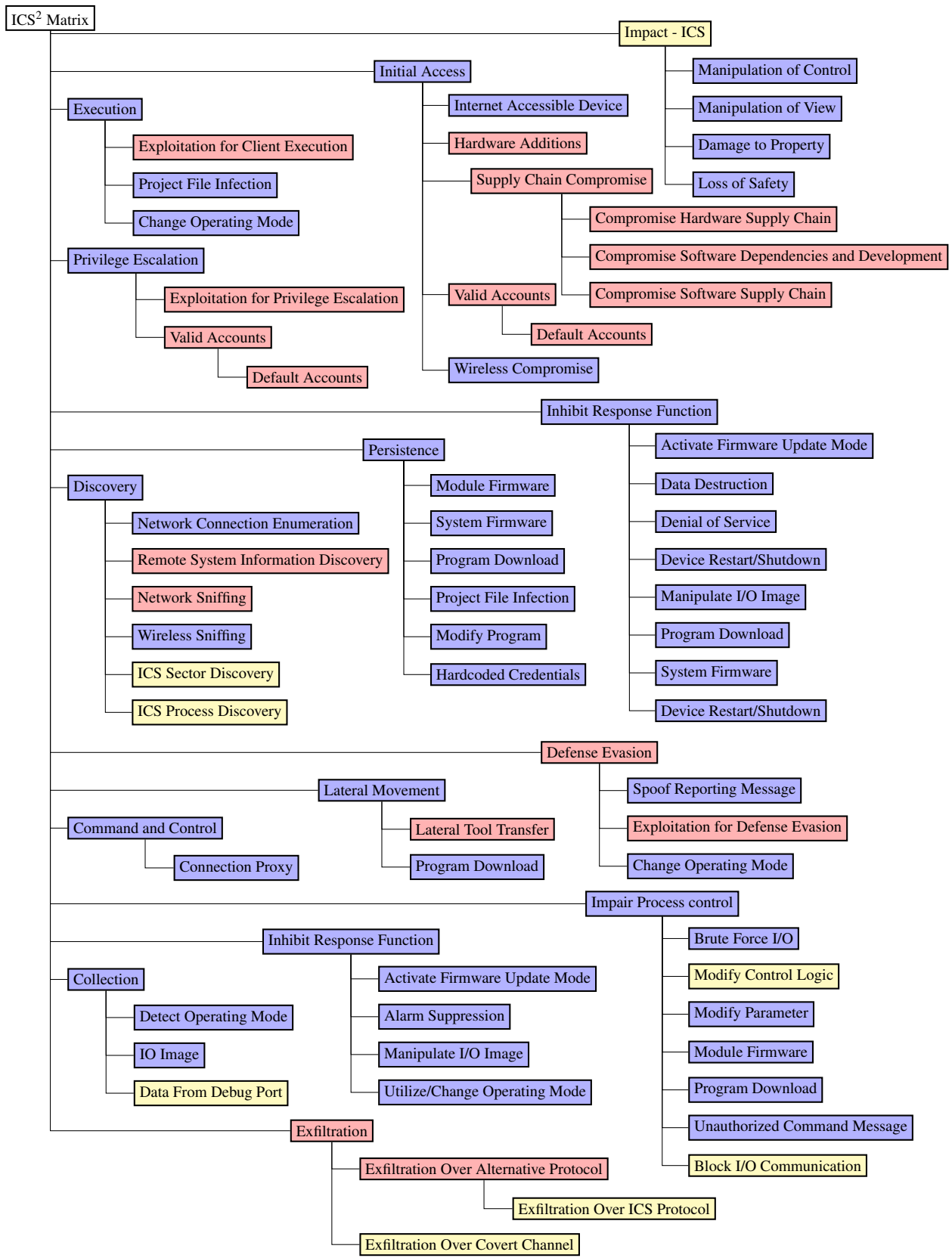

Figure 13: Condensed version of the ICS$^2$ Matrix. It includes nodes from the MITRE ATT&CK for ICS Matrix (Blue), the MITRE ATT&CK Enterprise (Red) and our additions (Yellow)